**Nanoscale Casimir force softening originated from quantum surface responses**

Hewan Zhang and Kun Ding†

*Department of Physics, State Key Laboratory of Surface Physics, and Key Laboratory of Micro and Nano Photonic Structures (Ministry of Education), Fudan University, Shanghai 200438, China*

† Corresponding E-mail: kunding@fudan.edu.cn

**Abstract**

Strong coupling between vacuum fields and quantum matter occurs at the nanoscale and broadens the horizon of light-matter interaction. Nanoscale Casimir force, as an exhibition of vacuum fields, inevitably experiences the influence of surface electron responses due to their quantum character, which are ignorable in micron Casimir force. Here, we develop a three-dimensional conformal map method to tackle typical experimental configurations with surface electron contributions to Casimir force purposely and delicately included. Based on this method, we reveal that quantum surface responses (QSRs) can either enhance or suppress the nanoscale Casimir force, depending on materials and crystal facets. The mechanism is demonstrated to be the Casimir force softening, which results from QSRs effectively altering the distance seen by the Casimir interaction. With such an understanding, we then provide a recipe to handle the nanoscale Casimir force between nanoscale complex objects. Our findings not only highlight the interaction between QSRs and vacuum fields but also provide a recipe for theoretical and experimental investigation of nanoscale fluctuation-type problems.

***Introduction***.—Vacuum fluctuations of electromagnetic fields are rooted in the uncertainty principle and thus ubiquitous, ineluctably letting vacuum fields or virtual photons have roles in the light-matter interaction [1,2]. Casimir force is one prominent example revealing the significance of vacuum fields in such interaction because Casimir energy between noncontact objects reflects the change of zero-point energy of electromagnetic fields influenced by the matter of objects [3-13]. Hence, various materials, especially emerging materials [14-25], have unraveled the features of their novel excitations in the Casimir force. More excitingly, recent experiments further disclose that vacuum fields have backactions on the matter [26-29], such as the breakdown of the quantum Hall effect [28]. All highlight the criticality of strong coupling between vacuum fields and excitations with quantum nature [27,30,31].

One of the prevailing approaches to such strong coupling is bringing objects in close proximity [32-34]. The left panel of Fig. 1(a) depicts two experimentally accomplishable prototypes, namely by attaching a sphere to an atomic force microscopy tip or using an optical trap [9,35-40]. The gap between objects in both setups can go below 20 nm, and Casimir force dressed by the novel matter excitations will become more significant than in the several hundred nanometers or above [right panel in Fig. 1(a)] by conventional calculations [41-44]. Besides excitations already included, strong coupling in conjunction with the scale certainly will kick previously ignorable effects in, even for some well-studied materials [32-34]. Take metal as an example, and quantum effects emerge firstly from the surface at the nanoscale (1~20 nm), dominantly coming from inhomogeneity of induced electron density and also from the kinetic energy of electrons [45-50], which begs for a method with all these quantum surface responses (QSRs) embedded to investigate nanoscale Casimir force between nanoscale objects.

The apparent approach is seemingly the ab initio calculations, but the algorithm for computing nanoscale Casimir forces with full electronic wavefunctions is few, even under the jellium model, and only for planar plates with thicknesses up to a few nanometers [51-53]. Starting from classical treatments, the hydrodynamic and non-local models have addressed complex surface effects on Casimir forces [54-57] but do not accurately consider QSRs, hindering the generalization to various realistic materials [58,59]. For accounting for the QSRs, the surface response function, which is known as Feibelman $d$-parameters in planar surfaces, should be considered [60]. Like bulk response functions $\varepsilon$ and $\mu$, $d$-parameters depend on the induced electron density and current density for different materials from ab initio calculations [50,58-62]. The superiority of $d$-parameters is that they faithfully reflect responses from surface electrons, such as the reference image plane treatments to van der Waals interaction between an atom and a metal surface [58,63,64]. Meanwhile, for effectively tackling nanoscale complex objects with relatively accurate surface responses, an excellent approach recently proposed is to incorporate $d$-parameters in the boundary conditions (BCs) of electromagnetic fields [65], which has been validated in a plethora of studies [66-72]. However, the impact of QSRs on nanoscale Casimir force between nanoscale objects has not yet been explored, which not only requires an accurate and efficient method to handle shapes and QSRs simultaneously but also begs for a theoretical framework to understand their roles.

In this work, we holistically investigate the nanoscale Casimir force of sphere-plate and bi-sphere systems shown in the top panel of Fig. 1(b), into which conceivable experimental setups in Fig. 1(a) are abstracted. To handle the interplay between shapes and QSRs, we develop a three-dimensional (3D) conformal map (CM) method incorporating QSRs because brute-force Casimir solvers are not that efficient (see Sec. I in Ref. [73]). We further demonstrate that QSRs for distinct metals and crystal planes can enhance or suppress nanoscale Casimir force compared to solely bulk contributions. To showcase the underlying mechanism, we utilize an analytical approach formulated under proximity force approximation (PFA) to disclose the nanoscale Casimir force softening effect, which delivers a recipe to investigate nanoscale Casimir force between nanoscale objects.

***Three-dimensional conformal map method.***—Sphere-plate and bi-sphere systems in physical space [top in Fig. 1(b)] can be transformed into an analytically tractable spherical shell system in auxiliary space [bottom in Fig. 1(b)] by 3D conformal inversion transformations $\boldsymbol{r}-\boldsymbol{r}_0=-R_T^2\,(\boldsymbol{R}-\boldsymbol{R}_0)/|\boldsymbol{R}-\boldsymbol{R}_0|^2$ [74,75], where $\boldsymbol{R}=(x,y,z)$ and $\boldsymbol{r}=(u,v,w)$ are coordinates in physical and auxiliary spaces, $\boldsymbol{R}_0$ (stars) and $\boldsymbol{r}_0$ (dots) denote inversion points, and $R_T$ is an arbitrary scaling length. The scale of interest indicates that the quasi-static approximation, i.e., $\boldsymbol{E}=-\boldsymbol{\nabla}\varphi$, can be adopted, and the impact of retardations will be discussed later. Together with the conformality of transformation, we only need to solve the Poisson equation $\boldsymbol{\nabla}\cdot\varepsilon(\boldsymbol{r})\boldsymbol{\nabla}\varphi(\boldsymbol{r})=0$ in the auxiliary space, where $\varepsilon(\boldsymbol{r})$ and $\varphi(\boldsymbol{r})$ are permittivity and electrostatic potential, respectively (see Sec. II in Ref. [73]).

The central point then becomes how to implement $d$-parameter BCs in the auxiliary space. For planar surfaces, two independent $d$-parameters, $d_\perp$ and $d_\parallel$, respectively, characterize the induced electron density ($n_1$) and parallel components of the induced current density [58]. In terms of BCs, $d_\perp$ and $d_\parallel$ characterize the response from surface dipoles and currents, and thus, $d_\sigma$ ($\sigma=\perp,\parallel$) must have the dimension of length with its typical values smaller than 1 nm. Figures 2(a) and 2(b) represent $d_\perp(\omega)$ for sodium (Na) and silver (Ag), which is reproduced from Refs. [50,69]. Since the nanoscale is beyond the scale of $d_\sigma$, the $d$-parameter BCs can then utilize $d_\sigma$ of planar surfaces [65]. Especially for our case, only the BCs involving $\boldsymbol{E}_{\parallel}^{\mathrm{phys}}$ and $D_{\perp}^{\mathrm{phys}}$ in the physical space should be considered. By employing the aforementioned CM, the BCs in the auxiliary space are $[\![\boldsymbol{E}_{\parallel}]\!] = -d_{\perp}\boldsymbol{\nabla}_{\parallel}[\![\Gamma E_{\perp}]\!]$ and $[\![D_{\perp}]\!] = d_{\parallel}\Gamma\boldsymbol{\nabla}_{\parallel}\cdot[\![\boldsymbol{D}_{\parallel}]\!]$, where $[\![\ldots]\!]$ denotes the difference of electromagnetic fields approaching the interface from both sides and $\Gamma = |\boldsymbol{r}-\boldsymbol{r}_0|^2/R_T^2$ (see Sec. III in Ref. [73]).

Since we only need to handle the Poisson equation, the BCs are further expressed as

$$[\![\varphi]\!] = -d_{\perp}\Gamma\nabla_{\perp}[\![\varphi]\!], \qquad [\![\varepsilon\nabla_{\perp}\varphi]\!] = d_{\parallel}\Gamma\boldsymbol{\nabla}_{\parallel}\cdot[\![\varepsilon\boldsymbol{\nabla}_{\parallel}\varphi]\!]. \tag{1}$$

By using a transformation $\varphi(\boldsymbol{r}) = |\boldsymbol{r}-\boldsymbol{r}_0|V(\boldsymbol{r})$ [83], the Poisson equation becomes a more readily solvable Laplace equation as $\boldsymbol{\nabla}^2 V(\boldsymbol{r}) = 0$. The general solution is expanded as $V(\boldsymbol{r}) = \sum_{l,m}\left[a_{l,m}^{+}\frac{r^l}{R_0^l} + a_{l,m}^{-}\frac{R_0^{l+1}}{r^{l+1}}\right]Y_l^m(\theta,\phi)$ in terms of regular and irregular solid harmonics in different regions. The expansion coefficients are $a_{l,m}^{\mathrm{in/out}}$, $a_{l,m}^{\pm}$, and $a_{l,m}^{\mathrm{s}\pm}$, with the superscripts representing the region and field property. By matching Eq. (1), we obtain the response matrix equation (see Sec. IV in Ref. [73])

$$\mathbf{S}_{m,\mathrm{wd}}\begin{pmatrix}\boldsymbol{a}_m^{+}\\ \boldsymbol{a}_m^{-}\end{pmatrix} = \begin{pmatrix}\boldsymbol{a}_m^{\mathrm{s}+}\\ \boldsymbol{a}_m^{\mathrm{s}-}\end{pmatrix}, \qquad \mathbf{S}_{m,\mathrm{wd}} = \begin{pmatrix}-\mathbf{I} & {\mathbf{S}_{m,\mathrm{wd}}^{++}}^{-1}\mathbf{T}_{m,\mathrm{wd}}^{+-}\\ {\mathbf{S}_{m,\mathrm{wd}}^{--}}^{-1}\mathbf{T}_{m,\mathrm{wd}}^{-+} & -\mathbf{I}\end{pmatrix}, \tag{2}$$

where $\boldsymbol{a}_m^{(\mathrm{s})\pm}$ is a column vector composed of $a_{l,m}^{(\mathrm{s})\pm}$ ($l = |m|, |m|+1, |m|+2, \ldots$) defined for the induced fields and source fields $\boldsymbol{E}_0$ and $\boldsymbol{p}$ [Fig. 1(b)]. $\mathbf{S}_{m,\mathrm{wd}}$ describes the system response (subscript denoting *d*-parameter BCs included), which is determined by geometry, bulk properties, and $d_\sigma$ (see Sec. V in Ref. [73] for explicit expressions). Numerically, we should specify a truncation $l = l_c$, making each matrix in Eq. (2) become the $(l_c - |m| + 1)\times(l_c - |m| + 1)$ matrix.

Before calculating the nanoscale Casimir force, we validate by comparing absorption spectra and field distributions with the finite element method performed for bi-sphere systems (see Sec. V in Ref. [73]). The retardation effect can be embedded as a correction in the 3D-CM method, which has also been verified by such comparison (see Sec. V in Ref. [73]). Note that the used *d*-parameters shall be consistent with bulk permittivity [50,69]. The fact that our 3D-CM method agrees with brute-force calculations but has a much lower computational cost indicates that it is an excellent candidate for investigating nanoscale Casimir force.

***Quantum surface response correction factor in Casimir force.***—By using $\mathbf{S}_{m,\mathrm{wd}}$, the Casimir energy of our system in Fig. 1(b) can then be evaluated in the imaginary frequency axis by the Lifshitz formula as [44]

$$E = \frac{\hbar}{4\pi} \sum_{m=0}^{+\infty} \int_{-\infty}^{+\infty} \ln f_m(i\xi)\, \mathrm{d}\xi\,, \tag{3}$$

where $f_m(\omega) = f_m^{\mathrm{c}} / f_m^{\mathrm{s}}$ is the mode condition function $f_m^{\mathrm{c}}$ for coupling bodies normalized by $f_m^{\mathrm{s}}$ for two single bodies (see details in Sec. VI, Ref. [73]). The Casimir force $F = -\partial E / \partial \delta$ of the sphere-plate system is exhibited in Figs. 2(c) and 2(d) for Na and Ag. Compared with the classical results [gray and black lines in Figs. 2(c) and 2(d)], the Casimir force with *d*-parameters is enhanced for Na [red lines in Fig. 2(c)], while suppressed for Ag whatever the crystal facets are [cyan and blue lines in Fig. 2(d)]. As gap size $\delta$ diminishes to the nanoscale, the contribution from *d*-parameters to Casimir force becomes considerable. To quantify such impact, we define the QSR correction factor as

$$\Xi_{\mathrm{QSR}} = \frac{F_{\mathrm{wd}}}{F_{\mathrm{cl}}} - 1, \tag{4}$$

where $F_{\mathrm{wd}}$ ($F_{\mathrm{cl}}$) denotes Casimir force with (without) *d*-parameters. The sign of $\Xi_{\mathrm{QSR}}$ indicates QSRs will enhance or suppress Casimir force, while its magnitude is relative change amounts of Casimir force. The solid lines in Fig. 3(b) show $\Xi_{\mathrm{QSR}}$ as a function of $\delta$ for results in Figs. 2(c) and 2(d). We see that QSRs lead to an increase of $1\%$ to $20\%$ in the nanoscale Casimir force for the Na case while a reduction of $0.4\%$ to $11\%$ for the Ag cases. Moreover, there is more substantial suppression in the Ag(111) case than in the Ag(100) case. Going beyond the nanoscale ($> 20\mathrm{nm}$), $\Xi_{\mathrm{QSR}}$ approaches zero, showing that QSR contributions to the Casimir force are insignificant. Hence, although such dramatic differences in nanoscale Casimir force for various metals and crystal facets can be qualitatively understood by distinct behaviors of QSRs, it is undoubtedly worth pursuing an analytical prescription based on the *d*-parameters, which will help to digest the role of QSRs in the Casimir force.

Recognizing the fact that geometric curvature is less crucial when $\delta \ll R_1$, we adopt the PFA treatment to sphere-plate systems, which is also one prevailing and conductive tool for experimentally interpreting the Casimir force [8,9,11,38,92,93]. As illustrated in Fig. 3(a), the spherical surface is discretized into annularly flat surfaces due to geometry. The Casimir force in the sphere-plate system is approximately the piecewise sum of the Casimir forces of every pairwise surface with local separations $L_i$. Utilizing identical *d*-parameters for all locally pairwise surfaces, we obtain the PFA results [inverted triangles in Fig. 3(b)], which are qualitatively suitable for all $\delta$ and quantitatively coincide with the 3D-CM results when $\delta \ll R_1$. The validity of PFA here actually separates the impact of geometry and QSRs on the nanoscale Casimir force for the system claimed in Figs. 1(a) and 1(b), making it possible to analyze the role of QSRs therein independently.

***Casimir softening in the reduction factor***.—To figure out how $d_\sigma$ affects Casimir force, we recall the reduction factor $\eta_F = F/F_{\mathrm{PEC}}$ defined for two parallel plates [right in Fig. 3(a)], which is the ratio of Casimir force between realistic materials $F$ and perfect electric conductors $F_{\mathrm{PEC}}$. At the nanoscale ($L \ll \lambda_p = 2\pi c/\omega_p$) and within the classical framework, the reduction factor is analytically shown as $\eta_{\mathrm{F}} = \alpha L/\lambda_p$ [87,88]. The coefficient $\alpha$ depends on the permittivity of metal, and $\alpha = 1.193$ for a non-dissipative Drude metal, i.e., $\epsilon_m = \epsilon_b - \omega_p^2/\omega(\omega + i\gamma)$ when $\epsilon_b = 1$ and $\gamma = 0$. By considering $d$-parameters, we analytically demonstrate that $\eta_F$ still reserves its form but with the physical distance term altered, namely (see Sec. VII in Ref. [73] for details)

$$\eta_F = \alpha \frac{L_\eta}{\lambda_p}, \qquad L_\eta = L \int_0^\infty \mathrm{d}K \int_0^\infty \mathrm{d}\Omega \, g[K, \Omega, d_\sigma(i\xi)], \tag{5}$$

where $K = qL$ and $\Omega = \xi/\omega_p$ are the normalized parallel wavenumber and imaginary frequency, and $\alpha = \frac{240}{\pi^3} \int_0^\infty \mathrm{d}K \int_0^\infty \mathrm{d}\Omega \left\{ e^{2K} \left( r_p^{\mathrm{wd}}[d_\sigma = 0] \right)^{-2} - 1 \right\}^{-1} K^2$. $L_\eta$ is a functional on $d_\sigma$, with $g = \frac{240}{\pi^3 \alpha} \left\{ e^{2K} \left( r_p^{\mathrm{wd}}[d_\sigma] \right)^{-2} - 1 \right\}^{-1} K^2$ being its spectral density, in which $r_p^{\mathrm{wd}}[d_\sigma] = \frac{(\epsilon_b - 1)\Omega(\Omega + \gamma_p) + 1 + K d_\perp/L + K d_\parallel/L}{(\epsilon_b + 1)\Omega(\Omega + \gamma_p) + 1 - K d_\perp/L + K d_\parallel/L}$ is the reflection coefficient of $p$-polarized waves [60,66]. When $d_\sigma = 0$, $L_\eta$ reverts to $L$. Nonzero $d_\sigma$ will lead to $L_\eta \neq L$, and $L_\eta > L$ ($L_\eta < L$) indicates that QSRs enhance (suppress) the Casimir force compared with the classical bulk one. Such distance modification due to details in the short-range shares the same spirit with that in the softening for gravitational force [94], and we thus dub $L_\eta/L$ as the Casimir force softening parameter in the reduction factor. Equation (5) has been validated by comparing it with the Lifshitz formula, so we plot $L_\eta/L$ calculated by Eq. (5) as a function of the physical distance in Fig. 3(c) by triangle markers. When $L$ is within (beyond) the nanoscale of interest here, $L_\eta$ deviates (approaches) $L$ as expected. For Na (Ag), $L_\eta/L$ is larger (smaller) than one, also verifying the observation in Figs. 2(c) and 2(d). The monotonic character of $L_\eta/L$ reveals that $d_\sigma$ softens the reduction factor incrementally as $L$ decreases, which is qualitatively scrutable but requires a more apparent comprehension.

Further, assuming that $d_\sigma$ is weakly dispersive and $d_\sigma \ll L$, we obtain an approximate analytical formula for $L_\eta$ as (see Sec. VII in Ref. [73] for details)

$$L_\eta(d_\sigma) = L + \sum_{\sigma=\perp,\parallel} C_\sigma d_\sigma(i\xi = 0)\,, \tag{6}$$

where the nondimensional coefficients $C_\sigma$ weight the $d_\sigma$ contributions to the Casimir force softening. The quantity $\sum C_\sigma d_\sigma$ visibly determines the nanoscale Casimir force by correction to distance. The results by Eq. (6) are plotted as dashed lines in Fig. 3(c), and good agreement with Eq. (5) for both metal and crystal facets is seen when $L$ is about $10$ nm. Minor deviation when $L < 3$ nm is due to the assumption $d_\sigma \ll L$, and thus, Eq. (6) offers a transparent way to analyze how $d_\sigma$ acts on $L_\eta$, where $C_\sigma$ becomes crucial then. $C_\sigma$ is determined by permittivity and positive definite, letting the sign of $d_\sigma(i\xi = 0)$ conduct enhancement or suppression of the nanoscale Casimir force. $d_\perp > 0$ $(< 0)$ characterizes the metal surface response with the spill-out (spill-in) effect, while $d_\parallel > 0$ $(< 0)$ indicates the excess (deficiency) of total electrons at the surface. Both positive (negative) situations intensify (diminish) the interactions between fluctuating charges and currents, consequently enhancing (suppressing) the Casimir force, as illustrated by Na (Ag). Concerning two crystal facets of Ag, $L_\eta$ of Ag(111) is lower than that of Ag(100) due to the values of $d_\perp(i\xi = 0)$ shown in Fig. 2(b), again verifying Eq. (6). Note that typical values of $C_\perp$ are larger than $C_\parallel$, indicating that the contribution to nanoscale Casimir force here is majorly from fluctuation dipoles but not fluctuation currents due to the scale and quasi-static approximation. The cruciality of the sign of $d_\sigma$ further highlights the Casimir softening recipe because $d_\sigma$ within the hydrodynamic and non-local models is always negative [54,89-91] (see Sec. VIII in Ref. [73] for details).

To further validate the Casimir softening mechanism, together with the fact that Casimir energy is the sum of zero-point energies, we also prove Eq. (6) from the analytical dispersion formula for parallel plates [59] (see Sec. VII in Ref. [73] for details). Moreover, the influence of retardation effects on the $d$-parameter corrections to the nanoscale Casimir force within the scale of interest here is minor (see Sec. IX in Ref. [73]).

***Discussions and conclusions***.—The Casimir softening treatments can be generalized to handle nanoscale objects with arbitrary shapes because Eq. (6) correlates to the reference surface position (see Sec. X in Ref. [73] for details). By simply shifting the small segments of surfaces of each object by $\sum C_\sigma d_\sigma/3$, the nanoscale Casimir forces are shown by dashed lines in Fig. 3(b), which agrees well with the 3D-CM results. To further reveal our recipe, we also calculate the Casimir force for the bi-sphere system, indicating that the Casimir softening framework developed above for the sphere-plate system is still valid in digesting the role of various metal and crystal facets in the bi-sphere system (see Sec. XI in Ref. [73]). From experimental measurement perspectives, the crucial issue is the asymptotic behavior of Casimir force when varying mutual distances. For the sphere-plate system, the Casimir force typically exhibits a power-law dependence on the gap size, namely $F \propto \delta^{n}$. Compared with classical results, nanoscale Casimir softening decreases (increases) the power-law exponent for the Na (Ag) case (see Sec. XII in Ref. [73]).

Successful implementation of the recipe from the sphere-plate system to the bi-sphere one delivers several messages for handling fluctuation-type problems in the nanoscale. Firstly, provided that the $d$-parameters are known, the Casimir softening framework is perhaps one of the most transparent and manageable approaches to investigating the Casimir force between nanostructures theoretically and experimentally. Due to the scale already experimentally accessible, our 3D-CM method not only makes up the gap between classical and fully quantum treatments of Casimir forces but also paves an alternative way to resolving the role of QSRs in other nanoscale fluctuation-type problems, such as Casimir torque [95,96], near-field thermal radiations [35,97-99], non-equilibrium thermal radiations [23,98,100], quantum sliding and rolling friction [101-105], and so on [106-108]. Last but not least, the crux of our method is to overcome the difficulty of complex surface behaviors and BCs by transforming them into an easy-handling space. Recognizing the development of computational conformal geometry [109], we believe our method can be generalized to handle a myriad of low-dimensional materials in complex geometry configurations.

In summary, by establishing a 3D-CM method to tackle geometry and QSRs simultaneously, we reveal that the nanoscale Casimir force can be enhanced or suppressed by QSRs unique for various metals and crystal facets. The underlying mechanism has been disclosed as the Casimir softening in the reduction factor, which originates from the fact that surface dipoles and currents effectively shift the boundary. Our findings not only underscore the significance of QSRs but also provide a recipe to theoretically and experimentally investigate the nanoscale Casimir force and even other fluctuation phenomena.

**Acknowledgment**

We thank Prof. Zhenghua An for the helpful discussions. This work is supported by the National Natural Science Foundation of China (12174072, 2021hwyq05), the National Key R&D Program of China (2022YFA1404701, 2022YFA1404500), and the Shanghai Science and Technology Innovation Action Plan (No. 24Z510205936).

## References

[1] P. W. Milonni, *The Quantum Vacuum: An Introduction to Quantum Electrodynamics* (Academic Press, San Diego, 1993).

[2] A. W. Rodriguez, F. Capasso, and S. G. Johnson, Nat. Photonics **5**, 211-221 (2011).

[3] H. B. G. Casimir, Proc. Kon. Ned. Akad. Wet. **51**, 793 (1948).

[4] S. K. Lamoreaux, Phys. Rev. Lett. **78**, 5 (1997).

[5] H. B. Chan, V. A. Aksyuk, R. N. Kleiman, D. J. Bishop, and F. Capasso, Science **291**, 1941-1944 (2001).

[6] G. Bressi, G. Carugno, R. Onofrio, and G. Ruoso, Phys. Rev. Lett. **88**, 041804 (2002).

[7] J. N. Munday, F. Capasso, and V. A. Parsegian, Nature **457**, 170-173 (2009).

[8] L. Tang, M. Wang, C. Y. Ng, M. Nikolic, C. T. Chan, A. W. Rodriguez, and H. B. Chan, Nat. Photonics **11**, 97-101 (2017).

[9] J. L. Garrett, D. A. T. Somers, and J. N. Munday, Phys. Rev. Lett. **120**, 040401 (2018).

[10] R. Zhao, L. Li, S. Yang, W. Bao, Y. Xia, P. Ashby, Y. Wang, and X. Zhang, Science **364**, 984–987 (2019).

[11] K. Y. Fong, H.-K. Li, R. Zhao, S. Yang, Y. Wang, and X. Zhang, Nature **576**, 243-247 (2019).

[12] K. Ding, D. Oue, C. T. Chan, and J. B. Pendry, Phys. Rev. Lett. **126**, 046802 (2021).

[13] T. Emig and G. Bimonte, Phys. Rev. Lett. **130**, 200401 (2023).

[14] F. S. S. Rosa, D. A. R. Dalvit, and P. W. Milonni, Phys. Rev. Lett. **100**, 183602 (2008).

[15] A. G. Grushin and A. Cortijo, Phys. Rev. Lett. **106**, 020403 (2011).

[16] V. Yannopapas and N. V. Vitanov, Phys. Rev. Lett. **103**, 120401 (2009).

[17] B. E. Sernelius, J. Phys. Condens. Matter **27**, 214017 (2015).

[18] J. H. Wilson, A. A. Allocca, and V. Galitski, Phys. Rev. B **91**, 235115 (2015).

[19] L. M. Woods, D. A. R. Dalvit, A. Tkatchenko, P. Rodriguez-Lopez, A. W. Rodriguez, and R. Podgornik, Rev. Mod. Phys. **88**, 045003 (2016).

[20] Y. Nishida, Phys. Rev. Lett. **130**, 096903 (2023).

[21] K. Nakata and K. Suzuki, Phys. Rev. Lett. **130**, 096702 (2023).

[22] D. Gelbwaser-Klimovsky, N. Graham, M. Kardar, and M. Kruger, Phys. Rev. Lett. **126**, 170401 (2021).

[23] Y. Tsurimaki, R. Yu, and S. Fan, Phys. Rev. B **107**, 115406 (2023).

[24] Y. Ema, M. Hazumi, H. Iizuka, K. Mukaida, and K. Nakayama, Phys. Rev. D **108**, 016009 (2023).

[25] K. Nakayama and K. Suzuki, arXiv:2310.18092 (2023).

[26] P. Forn-Díaz, L. Lamata, E. Rico, J. Kono, and E. Solano, Rev. Mod. Phys. **91**, 025005 (2019).

[27] F. J. Garcia-Vidal, C. Ciuti, and T. W. Ebbesen, Science **373**, 178 (2021).

[28] F. Appugliese, J. Enkner, G. L. Paravicini-Bagliani, M. Beck, C. Reichl, W. Wegscheider, G. Scalari, C. Ciuti, and J. Faist, Science **375**, 1030-1034 (2022).

[29] C. Roques-Carmes, Y. Salamin, J. Sloan, S. Choi, G. Velez, E. Koskas, N. Rivera, S. E. Kooi, J. D. Joannopoulos, and M. Soljačić, Science **381**, 205-209 (2023).

[30] N. Bartolo and C. Ciuti, Phys. Rev. B **98**, 205301 (2018).

[31] Y. Ke, Z. Song, and Q.-D. Jiang, Phys. Rev. Lett. **131**, 223601 (2023).

[32] J. A. Scholl, A. L. Koh, and J. A. Dionne, Nature **483**, 421-427 (2012).

[33] F. Benz, M. K. Schmidt, A. Dreismann, R. Chikkaraddy, Y. Zhang, A. Demetriadou, C. Carnegie, H. Ohadi, B. de Nijs, R. Esteban *et al.*, Science **354**, 726-729 (2016).

[34] R. Chikkaraddy, B. de Nijs, F. Benz, S. J. Barrow, O. A. Scherman, E. Rosta, A. Demetriadou, P. Fox, O. Hess, and J. J. Baumberg, Nature **535**, 127-130 (2016).

[35] K. Kloppstech, N. Könne, S.-A. Biehs, A. W. Rodriguez, L. Worbes, D. Hellmann, and A. Kittel, Nat. Commun. **8**, 14475 (2017).

[36] J. Ahn, Z. Xu, J. Bang, Y. H. Deng, T. M. Hoang, Q. Han, R. M. Ma, and T. Li, Phys. Rev. Lett. **121**, 033603 (2018).

[37] J. Ahn, Z. Xu, J. Bang, P. Ju, X. Gao, and T. Li, Nat. Nanotechnol. **15**, 89-93 (2020).

[38] Z. Xu, P. Ju, X. Gao, K. Shen, Z. Jacob, and T. Li, Nat. Commun. **13**, 6148 (2022).

[39] J. Vijayan, Z. Zhang, J. Piotrowski, D. Windey, F. van der Laan, M. Frimmer, and L. Novotny, Nat. Nanotechnol. **18**, 49-54 (2022).

[40] V. Liška, T. Zemánková, V. Svak, P. Jákl, J. Ježek, M. Bránecký, S. H. Simpson, P. Zemánek, and O. Brzobohatý, Optica **10**, 1203-1209 (2023).

[41] E. M. Lifshitz and L. P. Pitaevskii, *Statistical Physics: Theory of the Condensed State* (Butterworth-Heinemann Ltd, Oxford, 1995).

[42] W. Beiglböck, *Casimir Physics* (Springer, Heidelberg, 2011).

[43] J. N. Israelachvili, *Intermolecular and surface forces* (Elsevier, California, 2011).

[44] B. E. Sernelius, *Fundamentals of van der Waals and Casimir interactions* (Springer, Sweden, 2018).

[45] G. Toscano, J. Straubel, A. Kwiatkowski, C. Rockstuhl, F. Evers, H. Xu, N. A. Mortensen, and M. Wubs, Nat. Commun. **6**, 7132 (2015).

[46] W. Yan, Phys. Rev. B **91**, 115416 (2015).

[47] D. Jin, Q. Hu, D. Neuhauser, F. von Cube, Y. Yang, R. Sachan, T. S. Luk, D. C. Bell, and N. X. Fang, Phys. Rev. Lett. **115**, 193901 (2015).

[48] K. Ding and C. T. Chan, Phys. Rev. B **96**, 125134 (2017).

[49] M. K. Svendsen, C. Wolff, A. P. Jauho, N. A. Mortensen, and C. Tserkezis, J. Phys. Condens. Matter **32**, 395702 (2020).

[50] A. Rodríguez Echarri, P. A. D. Gonçalves, C. Tserkezis, F. J. García de Abajo, N. A. Mortensen, and J. D. Cox, Optica **8**, 710-721 (2021).

[51] V. Despoja, M. Šunjić, and L. Marušić, Phys. Rev. B **83**, 165421 (2011).

[52] L. Marušić, V. Despoja, and M. Šunjić, Solid State Commun. **151**, 1363-1366 (2011).

[53] H.-Y. Wu, J. Phys. Commun. **2**, 085005 (2018).

[54] R. Esquivel, C. Villarreal, and W. L. Mochán, Phys. Rev. A **68**, 052103 (2003).

[55] A. M. Contreras-Reyes and W. L. Mochán, Phys. Rev. A **72**, 034102 (2005).

[56] R. Esquivel-Sirvent, C. Villarreal, W. L. Mochán, A. M. Contreras-Reyes, and V. B. Svetovoy, J. Phys. A: Math. Gen. **39**, 6323-6331 (2006).

[57] Y. Luo, R. Zhao, and J. B. Pendry, Proc. Natl. Acad. Sci. U.S.A. **111**, 18422–18427 (2014).

[58] A. Liebsch, *Electronic Excitations at Metal Surfaces* (Springer Science & Business Media, New York, 1997).

[59] T. Christensen, W. Yan, A. P. Jauho, M. Soljacic, and N. A. Mortensen, Phys. Rev. Lett. **118**, 157402 (2017).

[60] P. J. Feibelman, Prog. Surf. Sci. **12**, 287-407 (1982).

[61] W. Yan, M. Wubs, and N. Asger Mortensen, Phys. Rev. Lett. **115**, 137403 (2015).

[62] S. Tanaka, T. Yoshida, K. Watanabe, Y. Matsumoto, T. Yasuike, M. Petrovic, and M. Kralj, Phys. Rev. Lett. **125**, 126802 (2020).

[63] B. N. J. Persson and P. Apell, Phys. Rev. B **27**, 6058-6065 (1983).

[64] B. N. J. Persson and E. Zaremba, Phys. Rev. B **30**, 5669-5679 (1984).

[65] Y. Yang, D. Zhu, W. Yan, A. Agarwal, M. Zheng, J. D. Joannopoulos, P. Lalanne, T. Christensen, K. K. Berggren, and M. Soljačić, Nature **576**, 248-252 (2019).

[66] P. A. D. Gonçalves, T. Christensen, N. Rivera, A.-P. Jauho, N. A. Mortensen, and M. Soljačić, Nat. Commun. **11**, 366 (2020).

[67] C. Tao, Y. Zhong, and H. Liu, Phys. Rev. Lett. **129**, 197401 (2022).

[68] Q. Zhou, P. Zhang, and X.-W. Chen, Phys. Rev. B **105**, 125419 (2022).

[69] F. Yang and K. Ding, Phys. Rev. B **105**, L121410 (2022).

[70] U. Hohenester and G. Unger, Phys. Rev. B **105**, 075428 (2022).

[71] P. A. D. Gonçalves and F. J. García de Abajo, Nano. Lett. **23**, 4242-4249 (2023).

[72] A. Babaze, T. Neuman, R. Esteban, J. Aizpurua, and A. G. Borisov, Nanophotonics **12**, 3277-3289 (2023).

[73] See Supplemental Material at the link for details of the 3D-CM and derivations of transformed BCs, response matrix equation, Lifshitz formula, Casimir softening, retardation effect, reference image surface method, and other numerical information, which includes Refs. [42,43,50,54,59,60,63-66,69,74-91].

[74] J. B. Pendry, A. I. Fernández-Domínguez, Y. Luo, and R. Zhao, Nat. Phys. **9**, 518-522 (2013).

[75] R. Zhao, Y. Luo, A. I. Fernandez-Dominguez, and J. B. Pendry, Phys. Rev. Lett. **111**, 033602 (2013).

[76] M. T. H. Reid, A. W. Rodriguez, J. White, and S. G. Johnson, Phys. Rev. Lett. **103**, 040401 (2009).

[77] D. E. Blair, *Inversion Theory and Conformal Mapping* (American Mathematical Society, U. S. A., 2000).

[78] U. Leonhardt and T. G. Philbin, in *Progress in Optics*, Vol. 53 (Elsevier, New York, 2009), pp. 69-152.

[79] N. B. Kundtz, D. R. Smith, and J. B. Pendry, Proc. IEEE **99**, 1622-1633 (2011).

[80] J. B. Pendry, D. Schurig, and D. R. Smith, Science **312**, 1780-1782 (2006).

[81] D. Schurig, J. J. Mock, B. J. Justice, S. A. Cummer, J. B. Pendry, A. F. Starr, and D. R. Smith, Science **314**, 977-980 (2006).

[82] M. Rahm, S. A. Cummer, D. Schurig, J. B. Pendry, and D. R. Smith, Phys. Rev. Lett. **100**, 063903 (2008).

[83] L. D. Landau, L. P. Pitaevskii, and E. M. Lifshitz, *Electrodynamics of Continuous Media* (Pergamon Press, Oxford, 1984).

[84] P. B. Johnson and R. W. Christy, Phys. Rev. B **6**, 4370-4379 (1972).

[85] E. D. Palik, *Handbook of Optical Constants of Solids II* (Academic, NewYork, 1991).

[86] M. Bordaga, U. Mohideenb, and V. M. Mostepanenko, Physics Reports **353**, 1 (2001).

[87] A. Lambrechta and S. Reynaud, Eur. Phys. J. Plus **8**, 309–318 (2000).

[88] F. Intravaia, C. Henkel, and A. Lambrecht, Phys. Rev. A **76**, 033820 (2007).

[89] R. Esquivel, C. Villarreal, and W. L. Mochán, Phys. Rev. A **71**, 029904(E) (2005).

[90] P. T. Kristensen, B. Beverungen, F. Intravaia, and K. Busch, Phys. Rev. B **108**, 205424 (2023).

[91] J. Sun, Y. Huang, and L. Gao, Phys. Rev. A **89**, 012508 (2014).

[92] B. V. Derjaguin, I. I. Abrikosova, and E. M. Lifshitz, Q. Rev. Chem. Soc. **10**, 295–329 (1956).

[93] A. A. Banishev, G. L. Klimchitskaya, V. M. Mostepanenko, and U. Mohideen, Phys. Rev. Lett. **110**, 137401 (2013).

[94] H. Das, S. Deb, and A. Baruah, Astrophys. J. **911**, 83 (2021).

[95] P. Thiyam, P. Parashar, K. V. Shajesh, O. I. Malyi, M. Boström, K. A. Milton, I. Brevik, and C. Persson, Phys. Rev. Lett. **120**, 131601 (2018).

[96] L. Chen and K. Chang, Phys. Rev. Lett. **125**, 047402 (2020).

[97] K. Kim, B. Song, V. Fernández-Hurtado, W. Lee, W. Jeong, L. Cui, D. Thompson, J. Feist, M. T. H. Reid, F. J. García-Vidal *et al.*, Nature **528**, 387-391 (2015).

[98] T. Zhu and J.-S. Wang, Phys. Rev. B **104**, L121409 (2021).

[99] G. Tang and J.-S. Wang, arXiv:2310.05417.

[100] R. Yu and S. Fan, Phys. Rev. Lett. **130**, 096902 (2023).

[101] J. B. Pendry, J. Phys. Condens. Matter **9**, 10301 (1997).

[102] A. Manjavacas and F. J. Garcia de Abajo, Phys. Rev. Lett. **105**, 113601 (2010).

[103] V. Despoja, P. M. Echenique, and M. Šunjić, Phys. Rev. B **83**, 205424 (2011).

[104] R. Zhao, A. Manjavacas, F. J. Garcia de Abajo, and J. B. Pendry, Phys. Rev. Lett. **109**, 123604 (2012).

[105] F. Intravaia, M. Oelschlager, D. Reiche, D. A. R. Dalvit, and K. Busch, Phys. Rev. Lett. **123**, 120401 (2019).

[106] C. R. Otey, W. T. Lau, and S. Fan, Phys. Rev. Lett. **104**, 154301 (2010).

[107] A. P. Raman, M. A. Anoma, L. Zhu, E. Rephaeli, and S. Fan, Nature **515**, 540-544 (2014).

[108] Y.-M. Zhang, T. Zhu, Z.-Q. Zhang, and J.-S. Wang, Phys. Rev. B **105**, 205421 (2022).

[109] X. D. Gu and S.-T. Yau, *Computational conformal geometry* (Higher Education Press & International Press, Beijing, 2008).

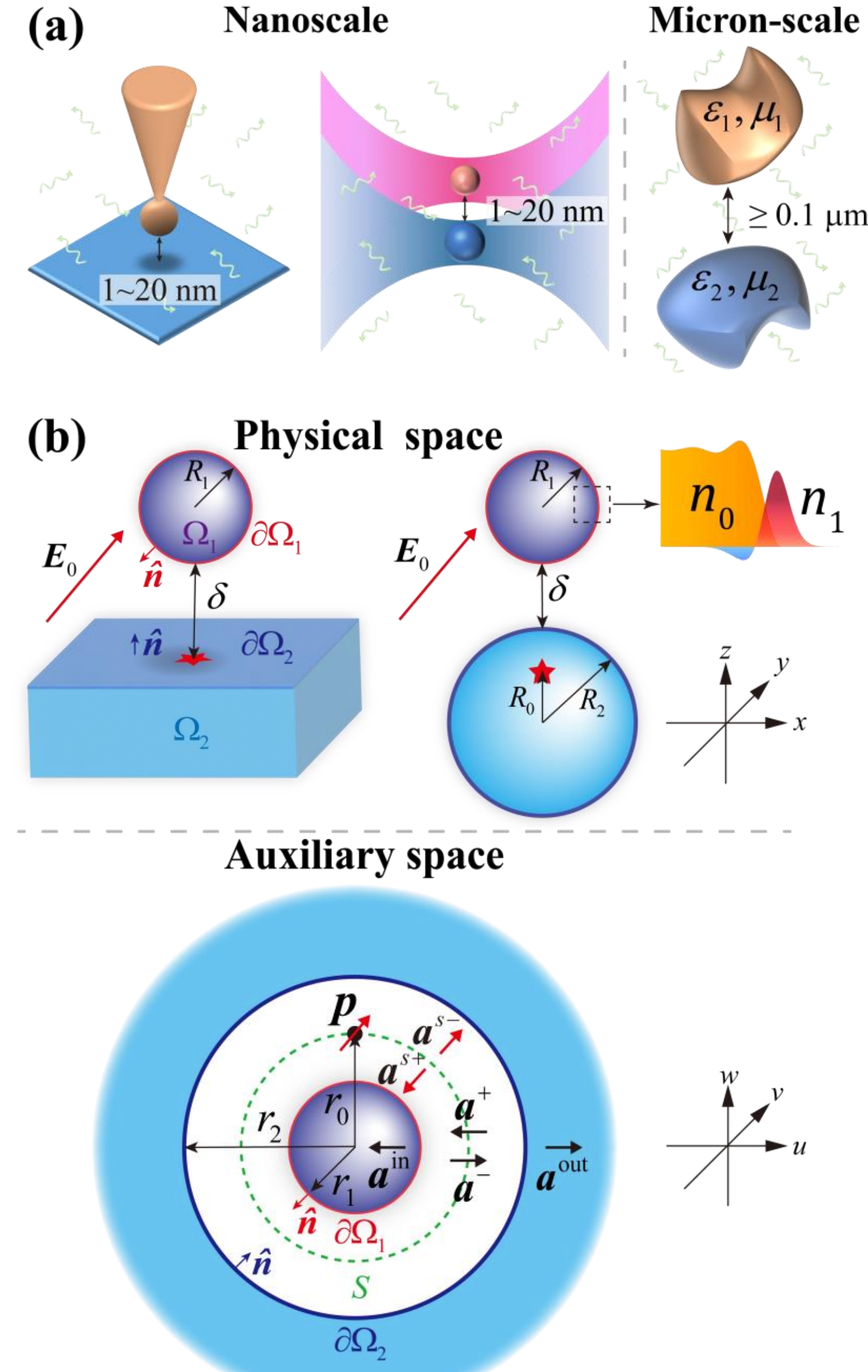

FIG. 1. (a) Schematic experimental setups possibly involving the nanoscale Casimir force (left panel). The left (right) panel shows the tip-substrate and bi-sphere system (two microscale objects with arbitrary shapes). The light green wiggles denote vacuum fluctuation. (b) The sphere-plate and bi-sphere systems in the physical space (upper panel) are mapped to a spherical shell system in the auxiliary space (lower panel) by a 3D CM. All geometric and physical definitions are illustrated in the figure. $\hat{\boldsymbol{n}}$ are the normal basis vectors to the surface $\partial\Omega_{1,2}$ in both spaces. The inset in the top right diagrammatically represents the distribution of $n_0$ and $n_1$ near the interface.

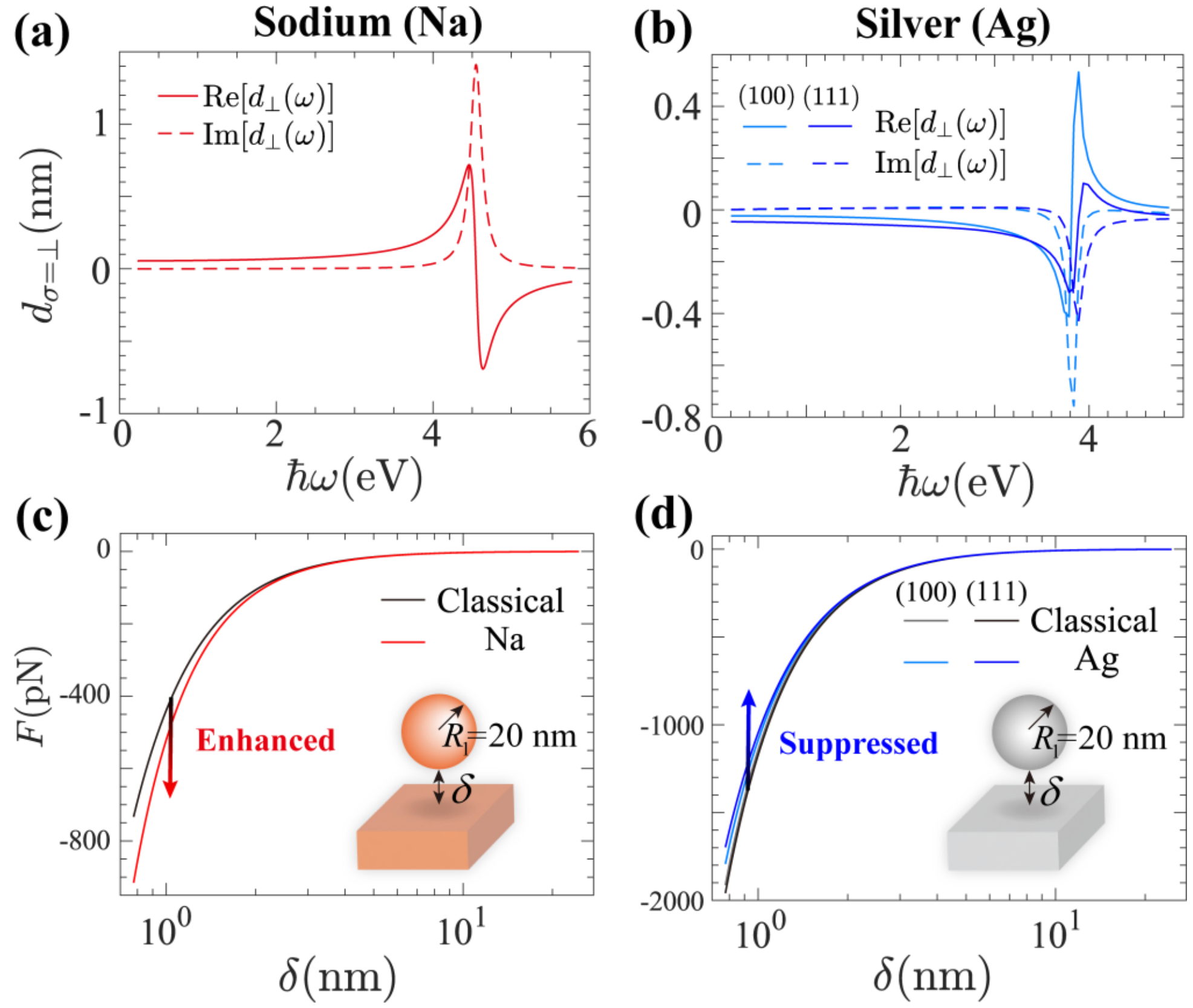

FIG. 2. (a-b) $d_\perp$ of a planar interface formed by vacuum and Na (a) or Ag (b). Re$[d_\perp(\omega)]$ and Im$[d_\perp(\omega)]$ are depicted by the solid and dashed lines, respectively. The lines in cyan and blue denote Ag(100) and Ag(111) surfaces. (c-d) Casimir force of the sphere-plate system as a function of $\delta$ for the Na (c) and Ag (d) cases. The black lines are the classical results ($d_\perp = d_\parallel = 0$), while the red, cyan, and blue lines show the Na, Ag(100), and Ag(111) results by using $d_\perp$ in (a-b). The explicit expression of $d_\perp(\omega)$ and the permittivity of both metals are shown in Sec. V, Ref. [73]. Other parameters are $R_1 = 20$ nm and $d_{\sigma=\parallel} = 0$.

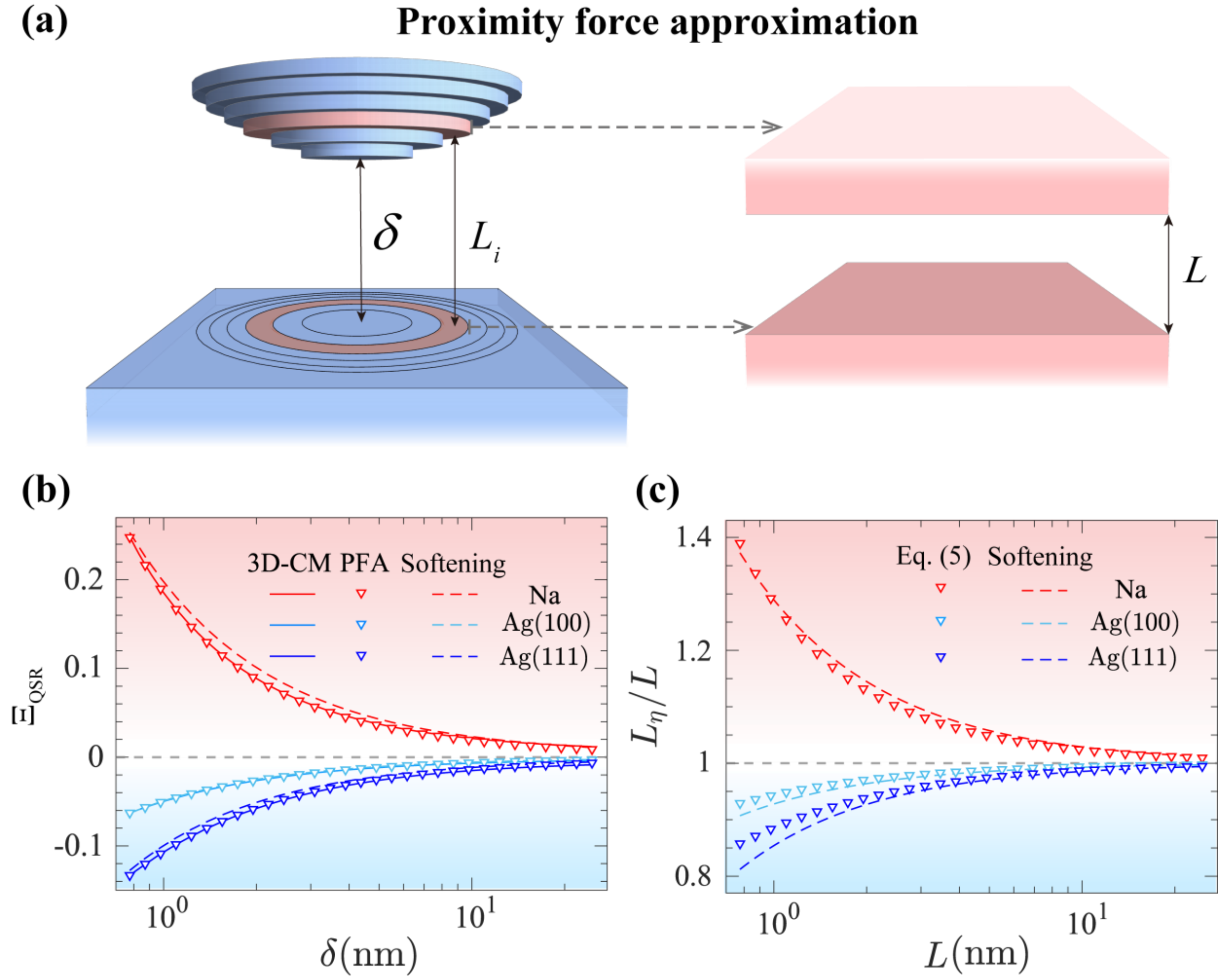

FIG. 3. (a) Schematics depicting PFA for the sphere-plate configuration (left panel). Each pairwise surface separated by $L_i$ is described by two semi-infinite parallel plates with the same distance (right panel). (b) $\Xi_{\mathrm{QSR}}$ for Casimir force in sphere-plate configuration. The solid lines, inverted triangle markers, and dashed lines represent the 3D-CM, PFA, and Casimir softening results, respectively. (c) $L_\eta/L$ as a function of physical distance $L$. The inverted triangle markers and dashed lines show fully integral [Eq. (5)] and Casimir softening results. The Na, Ag(100), and Ag(111) cases in (b-c) are in red, cyan, and blue, respectively.

## Supplemental Materials for
## Nanoscale Casimir force softening originated from quantum surface responses

Hewan Zhang and Kun Ding†

*Department of Physics, State Key Laboratory of Surface Physics, and Key Laboratory of Micro and Nano Photonic Structures (Ministry of Education), Fudan University, Shanghai 200438, China*

† Corresponding E-mail: kunding@fudan.edu.cn

## Section I. Comparisons of the SCUFF-EM solver and the conformal map

The problem we aim to address is to make clear the impacts of quantum surface responses (QSRs) of metals on the Casimir force for complex structures in nanoscale regimes [left panel in Fig. 1(a)], which is an unexplored subject. However, no computational Casimir solvers are employed to handle Maxwell equations and QSRs simultaneously. To address this problem, we have developed a three-dimensional (3D) conformal map (CM) method to handle QSRs and complex geometry simultaneously and accurately.

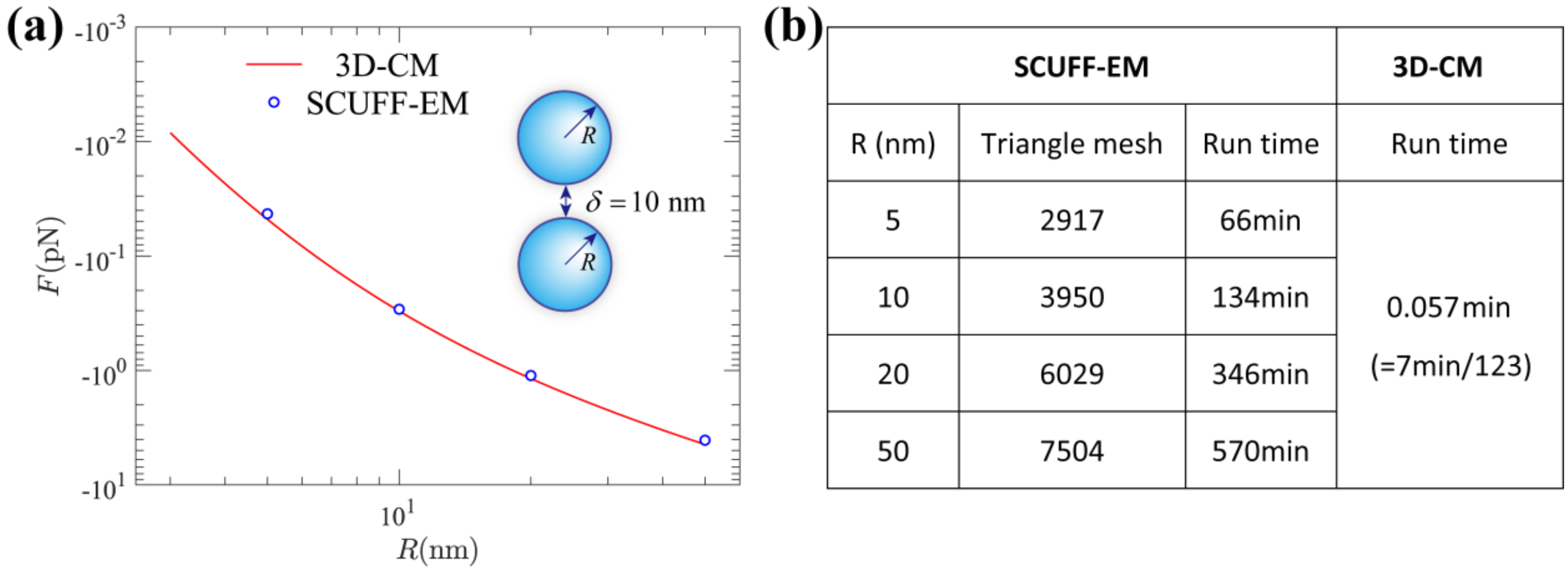

| SCUFF-EM | | | 3D-CM |
|---|---|---|---|
| R (nm) | Triangle mesh | Run time | Run time |
| 5 | 2917 | 66min | 0.057min (=7min/123) |
| 10 | 3950 | 134min | |
| 20 | 6029 | 346min | |
| 50 | 7504 | 570min | |

**Figure S1.** (a) Classical Casimir force as a function of sphere radius $R$. The gap size $\delta$ is 10 nm, as indicated by the schematic in the inset. The red line is from the 3D-CM method, while the blue circles are from SCUFF-EM. (b) shows the comparison of computational time between SCUFF-EM and 3D-CM. The number of triangle mesh and run time in the SCUFF-EM calculation [circles in (a)] using 192 CPU cores are depicted, while the total run time of the 3D-CM calculation [red line in (a)] for 123 $R$-points using 96 CPU cores are also shown. The used processor is “Intel(R) Xeon(R) Platinum 8469C”.

Meanwhile, the analyticity nature of 3D-CM spontaneously implies high efficiency [1,2], although there are some known Casimir solvers for micron-scale objects (using classical $\varepsilon$ and $\mu$) [right panel in Fig. 1(a)], such as boundary-element-method based solver (SCUFF-EM) [3]. However, this open-source package can now only compute the Casimir effect without QSRs, and even within such classical frameworks, the computational time cost is incredibly demanding. Figure S1 shows the results and the according computational time from SCUFF-EM and 3D-CM. The classical Casimir forces calculated by the 3D-CM method [red line in Fig. S1(a)] show good agreement with those obtained from SCUFF-EM [circles in Fig. S1(a)] but at a much lower computational time [column “Run time” in Fig. S1(b)]. The run time of SCUFF-EM for a single data point is at least 1 hour by using 192 CPU cores, while our 3D-CM code is less than 5 seconds. This highlights the merits of 3D-CM, i.e., within the scale we are interested in, the 3D-CM method shows excellent accuracy and efficiency.

**Section II. Details of the conformal map method**

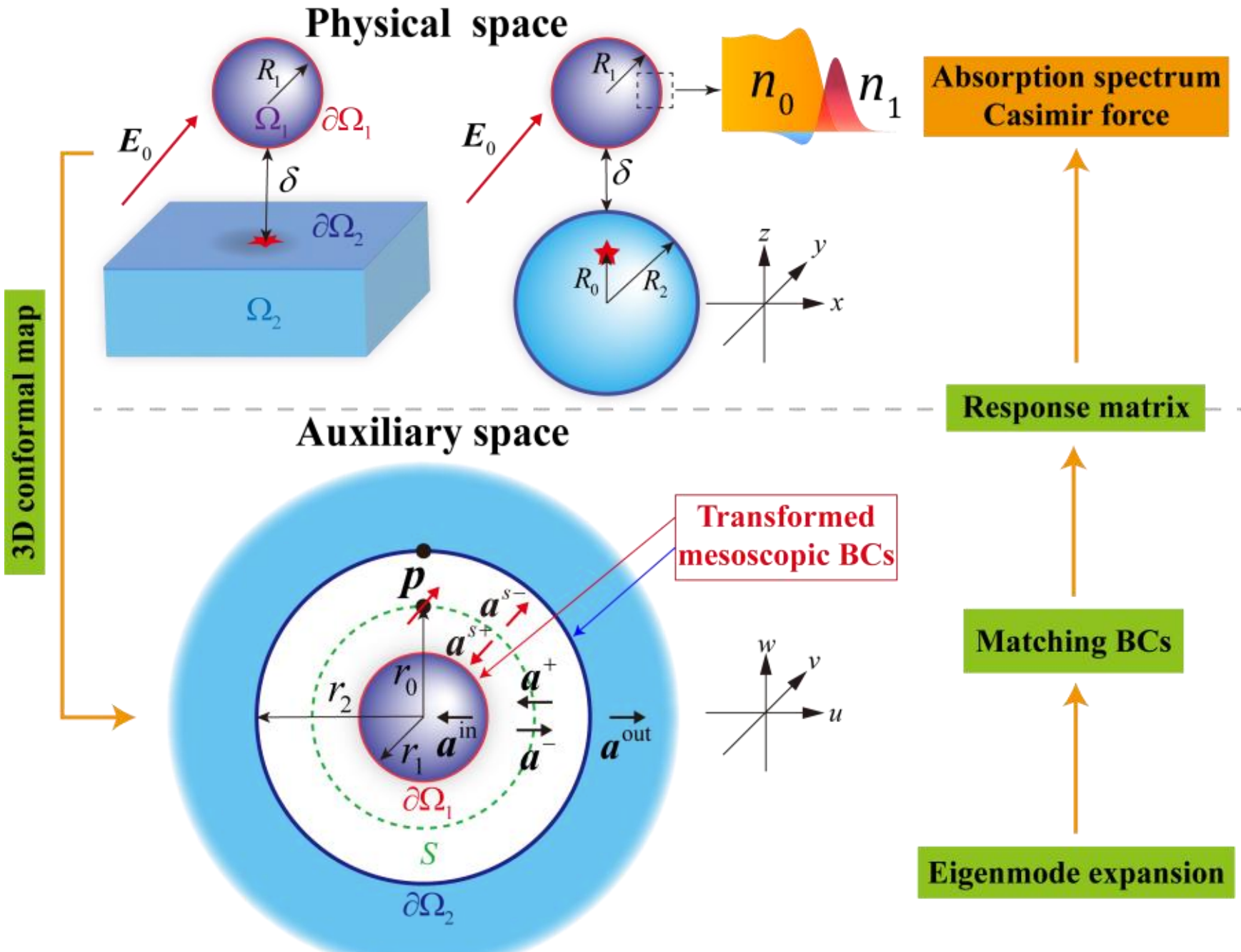

**Figure S2**. Flowchart of the 3D-CM method to the sphere-plate and bi-sphere systems. The main body is reproduced from Fig. 1(b) but with attached explanations of the method. The left-hand-side arrow shows the CM, while the right-hand-side panel depicts the logical flow of our method. On the top (orange box) are the aimed physical quantities, and at the bottom are several steps to acquire the targets (green boxes).

Figure S2 exhibits the flowchart of the 3D-CM method for solving physical systems. Firstly, we employ 3D-CM to transform the sphere-plate and bi-sphere systems in the physical space (upper panel) to the analytically tractable spherical shell system in the auxiliary space (lower panel). The top right schematically shows the conduction electron density $n_0$ at equilibrium (orange color), which is not evenly distributed inside the metals but spills over the jellium (macroscopic) boundary. Meanwhile, the induced electron density $n_1$ (red and blue color) does not concentrate on the surface as a delta function but actually spills across the interface. The Feibelman $d$-parameters are then introduced to describe the quantum behavior of electrons near the interface. The surface quantum effects can be integrated into mesoscopic boundary conditions (BCs) by $d$-parameters [4]. In the scale of interest here, surface responses of curved metal surfaces are assumed to be uniform, meaning that geometric corrugations become locally flat in the sense of QSRs, and thereby, the surface response functions $d_\sigma$ ($\sigma = \perp, \parallel$) at the planar metal-vacuum interface are applied. Hence, the crucial aspect of 3D CM lies in deriving the transformed mesoscopic BCs in the auxiliary space. Secondly, we solve the Laplace equation with respect to $V(\boldsymbol{r})$ in the auxiliary space. $V(\boldsymbol{r})$ is expanded in terms of eigenmodes, i.e., regular and irregular solid harmonics. $V(\boldsymbol{r})$ in different regions is matched through transformed mesoscopic BCs. Thirdly, after transforming back to physical space, the response matrix of physical systems can be obtained. Fourthly, the response matrix is employed for numerical computations, encompassing the absorption spectrum and Casimir force.

We detail the 3D CM in the following due to its central role here. According to Liouville's theorem [5], inversion transformation is a CM in Euclidean space capable of mapping spheres and planes into spheres and planes. Thus, the sphere-plate system and bi-sphere system in physical space can be mapped to a spherical shell system in the auxiliary space by the inversion transformation with proper inversion points [1], as exhibited in Fig. S2. Here, we choose the center of the light blue sphere as the origin of both $(x, y, z)$ and $(u, v, w)$. The inversion points marked by red pentagrams (black solid dots) for the bi-sphere and sphere-plate (spherical shell) systems are located at $\boldsymbol{R}_0 = (0,0,R_0)$ and $\boldsymbol{r}_0 = (0,0,r_0)$. The mapping relations between $\boldsymbol{R}$ in the physical space and $\boldsymbol{r}$ in the auxiliary space are [1]

$$\boldsymbol{r} - \boldsymbol{r}_0 = -R_T^2 \frac{\boldsymbol{R} - \boldsymbol{R}_0}{|\boldsymbol{R} - \boldsymbol{R}_0|^2}, \tag{S2.1}$$

$$\boldsymbol{R} - \boldsymbol{R}_0 = -R_T^2 \frac{\boldsymbol{r} - \boldsymbol{r}_0}{|\boldsymbol{r} - \boldsymbol{r}_0|^2}, \tag{S2.2}$$

where $R_T$ is an arbitrary scaling length. Domains and boundaries of the same colors are consistently mapped, with the purple domain and the corresponding red boundary denoted as $\Omega_1$ and $\partial\Omega_1$, and with the light blue domain and the corresponding deep blue boundary denoted as $\Omega_2$ and $\partial\Omega_2$. Moreover, we denote that the sphere radii of bi-spheres are $R_1$ and $R_2$, and the gap size is $\delta$, while the radii of the transformed spherical shells are $r_1$ and $r_2$. The flat plate $\Omega_2$ in the sphere-plate system is equivalent to the sphere with $R_2 = \infty$, and the inversion point is on $\partial\Omega_2$, marked by a solid dot, i.e., $R_0 = R_2$ and $r_0 = r_2$.

We now express geometric parameters in the auxiliary space in terms of those in the physical space. For the mapping of arbitrary points in $\partial\Omega_{1,2}$, we find

$$\frac{r_1}{r_0} = \frac{R_1}{h - R_0}, \qquad R_T^2 = (h - R_0) r_0 \left(1 - \frac{r_1^2}{r_0^2}\right), \tag{S2.3}$$

$$\frac{r_0}{r_2} = \frac{R_0}{R_2}, \qquad R_T^2 = r_0 R_0 \left(\frac{r_2^2}{r_0^2} - 1\right), \tag{S2.4}$$

where $h = R_1 + R_2 + \delta$ is the distance between the centers of two spheres. Remarkably, Eqs. (S2.3)-(S2.4) can be easily grasped from the mapping from $\boldsymbol{R} = (0,0,R_2+\delta), (0,0,h+R_1)$ to $\boldsymbol{r} = (0,0,-r_1), (0,0,r_1)$ and the mapping from $\boldsymbol{R} = (0,0,R_2), (0,0,-R_2)$ to $\boldsymbol{r} = (0,0,-r_2), (0,0,r_2)$. Defining geometric parameters $\chi_+ = \frac{r_1}{r_0}$, $\chi_- = \frac{r_0}{r_2}$, and $L_0 = \frac{R_T^2}{r_0}$ in the auxiliary space, Eqs. (S2.3)-(S2.4) can be recast as

$$\chi_+ = \frac{R_1}{h - R_0}, \qquad \chi_- = \frac{R_0}{R_2}, \qquad L_0 = (h - R_0)(1 - \chi_+^2), \tag{S2.5}$$

where $R_0 = \frac{h^2 - R_1^2 + R_2^2}{2h} - \sqrt{\left(\frac{h^2 - R_1^2 + R_2^2}{2h}\right)^2 - R_2^2}$. For the sphere-plate system, $R_0 = R_2 = +\infty$, Eq. (S2.5) reduces to

$$\chi_+ = 1 + \frac{\delta}{R_1} - \sqrt{2\frac{\delta}{R_1} + \left(\frac{\delta}{R_1}\right)^2}, \qquad \chi_- = 1, \qquad L_0 = 2\sqrt{2R_1\delta + \delta^2}. \tag{S2.6}$$

Notably, $R_0 = R_2$ indicates that the inversion point is located at the interface $\partial\Omega_1$, and now the origin of the coordinate in the physical space is situated at infinity.

Hypothetically, there exists an incident plane wave $\boldsymbol{E}_0$ in physical space, as shown in Fig. S2. Under the 3D CM [Eqs. (S2.1)-(S2.2)], $\boldsymbol{E}_0$ is mapped to an electric dipole $\boldsymbol{p}_0$ located at $\boldsymbol{r}_0$ in the auxiliary space under the quasi-static approximation [1], as indicated by a red arrow. In the auxiliary space, the concentric spherical surface, where the inversion point is located, is denoted as $S$, as marked by the green dashed line. For the spherical shell system mapped from the sphere-plate system, the surface $S$ overlays with $\partial\Omega_2$. The static electric potential of $\boldsymbol{p}_0$ can be expanded in terms of inward and outward propagating spherical waves, $\boldsymbol{a}^{s+}$ and $\boldsymbol{a}^{s-}$, inside and outside $S$.

## Section III. Transformed mesoscopic boundary conditions

In classical electrodynamics [6,7], if a transformation is conformal, macroscopic BCs are invariant, and so is the 3D CM defined by Eqs. (S2.1)-(S2.2). However, the mesoscopic BCs in the auxiliary space do not keep the same form as those in the physical space and, thus, are required to be derived. The Jacobian of Eqs. (S2.1)-(S2.2) is

$$\Lambda^i{}_\sigma = \frac{\partial x^i}{\partial x^\sigma}, \qquad \Lambda^\sigma{}_i = \frac{\partial x^\sigma}{\partial x^i}, \tag{S3.1}$$

where $i = x, y, z$ and $\sigma = u, v, w$. We define indices with English (Greek) letters for physical (auxiliary) space. Hence, the Jacobian determinants and the metric tensors are

$$\det \Lambda^{\sigma}{}_{i} = \Gamma^{3}, \qquad \det \Lambda^{i}{}_{\sigma} = \Gamma^{-3}, \tag{S3.2}$$

$$g^{ij} = g_{ij} = \mathbf{I}, \qquad g^{\sigma\rho} = \Lambda^{\sigma}{}_{i} g^{ij} \Lambda^{\rho}{}_{j} = \Gamma^{2}\mathbf{I}, \qquad g_{\sigma\rho} = \Gamma^{-2}\mathbf{I}, \tag{S3.3}$$

where $\Gamma = \frac{|\boldsymbol{r}-\boldsymbol{r}_0|^2}{R_T^2}$. For applying the coordinate transformation technique to solve electromagnetic problems, a crucial requirement is to preserve the form invariance of Maxwell's equations under the transformation [7-10]. The transformation rules of the differential operator $\boldsymbol{\nabla}$ in the Maxwell's equations are $\partial_i = \Lambda^{\sigma}{}_{i}\partial_{\sigma}$. To maintain this invariance, the electromagnetic fields in physical and auxiliary spaces are required to satisfy the transformation rules $E_i = \Lambda^{\sigma}{}_{i} E_{\sigma}$ and $H_i = \Lambda^{\sigma}{}_{i} H_{\sigma}$, while the constitutive parameters $\boldsymbol{\varepsilon}$ and $\boldsymbol{\mu}$ transform as $\varepsilon^{\sigma\rho} = \frac{\Lambda^{\sigma}{}_{i}\varepsilon^{ij}\Lambda^{\rho}{}_{j}}{\det \Lambda^{\sigma}{}_{i}}$ and $\mu^{\sigma\rho} = \frac{\Lambda^{\sigma}{}_{i}\mu^{ij}\Lambda^{\rho}{}_{j}}{\det \Lambda^{\sigma}{}_{i}}$ [7]. Employing these fundamental transformation rules along with the constitutive relations $\boldsymbol{D} = \boldsymbol{\varepsilon}\boldsymbol{E}$ and $\boldsymbol{B} = \boldsymbol{\mu}\boldsymbol{H}$, we obtain the transformation rules for another two electromagnetic quantities as $D^i = \frac{1}{\det \Lambda^{i}{}_{\sigma}} \Lambda^{i}{}_{\sigma} D^{\sigma}$ and $B^i = \frac{1}{\det \Lambda^{i}{}_{\sigma}} \Lambda^{i}{}_{\sigma} B^{\sigma}$. Furthermore, assuming that the permittivity and permeability in physical space are isotropic, i.e., $\varepsilon^{ij} = \varepsilon^{\text{phys}}\delta^{ij}$ and $\mu^{ij} = \mu^{\text{phys}}\delta^{ij}$, the corresponding quantities in auxiliary space are given by $\varepsilon^{\sigma\rho} = \Gamma^{-1}\varepsilon^{\text{phys}}\mathbf{I}$ and $\mu^{\sigma\rho} = \Gamma^{-1}\mu^{\text{phys}}\mathbf{I}$, which remain isotropic. Due to the conformality of the transformation, the $i$ or $\sigma$ components of the tangential and normal electromagnetic fields still follow the transformation rules of complete electromagnetic fields, i.e.,

$$E_{p,i} = \Lambda^{\sigma}{}_{i} E_{p,\sigma}, \tag{S3.4a}$$

$$H_{p,i} = \Lambda^{\sigma}{}_{i} H_{p,\sigma}, \tag{S3.4b}$$

$$D_p^i = \frac{1}{\det \Lambda^{i}{}_{\sigma}} \Lambda^{i}{}_{\sigma} D_p^{\sigma}, \tag{S3.4c}$$

$$B_p^i = \frac{1}{\det \Lambda^{i}{}_{\sigma}} \Lambda^{i}{}_{\sigma} B_p^{\sigma}, \tag{S3.4d}$$

where $p = \perp, \parallel$. Notably, the isotropy of constitutive parameters in both spaces, together with the conformality of the transformation, guarantees that the normal (tangential) components of $\boldsymbol{D}$ and $\boldsymbol{B}$ in the physical space are mapped exclusively into the normal (tangential) components in the auxiliary space, without mixing into other components. Moreover, the differential operators in the tangential and normal directions also satisfy the same transformation rules as the full differential operator

$$\partial_{p,i} = \Lambda^{\sigma}{}_{i}\partial_{p,\sigma}, \tag{S3.5}$$

where $\boldsymbol{\partial}_{\perp} = \boldsymbol{\nabla}_{\perp} = \hat{\boldsymbol{n}}\hat{\boldsymbol{n}} \cdot \boldsymbol{\nabla}$ and $\boldsymbol{\partial}_{\parallel} = \boldsymbol{\nabla}_{\parallel} = (\mathbf{I} - \hat{\boldsymbol{n}}\hat{\boldsymbol{n}}) \cdot \boldsymbol{\nabla}$. The normal basis vector $\hat{\boldsymbol{n}}^{\text{phys}}$ of the interface in physical space also follows a similar transformation

$$\hat{n}_i = \frac{1}{\Gamma}\Lambda^{\sigma}{}_{i}\hat{n}_{\sigma}, \tag{S3.6}$$

where the factor $\Gamma^{-1}$ arises from the requirement that $\hat{\boldsymbol{n}}^{\text{phys}} \cdot \hat{\boldsymbol{n}}^{\text{phys}} = 1$ and $\hat{\boldsymbol{n}} \cdot \hat{\boldsymbol{n}} = 1$. It is worth noting that $\left(E_{i,\sigma}, H_{i,\sigma}, D^{i,\sigma}, B^{i,\sigma}\right)$ are realistic electromagnetic field components, and $\hat{n}_{i,\sigma}$ is a realistic normal basis function. The conformality ensures that the electromagnetic fields and normal direction preserve angles before and after the transformation. Combining Eqs. (S3.4) and (S3.6), we find that the normal components of the electromagnetic field in the auxiliary space are

$$E_{\perp}^{\text{phys}} = \boldsymbol{E}^{\text{phys}} \cdot \hat{\boldsymbol{n}} = E_i n_j g^{ij} = \frac{1}{\Gamma} E_{\sigma} n_{\rho} g^{\sigma\rho} = \Gamma E_{\perp}, \tag{S3.7a}$$

$$H_{\perp}^{\text{phys}} = \Gamma H_{\perp}, \tag{S3.7b}$$

$$D_{\perp}^{\text{phys}} = \boldsymbol{D}^{\text{phys}} \cdot \hat{\boldsymbol{n}} = D^i n_i = \frac{1}{\Gamma \det \Lambda^{i}{}_{\sigma}} D^{\sigma} n_{\sigma} = \Gamma^2 D_{\perp}, \tag{S3.7c}$$

$$B_{\perp}^{\text{phys}} = \Gamma^2 B_{\perp}. \tag{S3.7d}$$

Equations (S3.7a)-(S3.7d) indicate that the conformality leads to a linear map between normal components in both spaces. So far, we have provided the necessary key ingredients to derive the transformation results for BCs. The mesoscopic BCs in physical space are

$$\left[\!\left[\boldsymbol{E}_{\parallel}^{\text{phys}}\right]\!\right] = -d_{\perp}\boldsymbol{\nabla}_{\parallel}^{\text{phys}}\left[\!\left[E_{\perp}^{\text{phys}}\right]\!\right], \tag{S3.8a}$$

$$\left[\!\left[\boldsymbol{H}_{\parallel}^{\text{phys}}\right]\!\right] = i\omega d_{\parallel}\left[\!\left[\boldsymbol{D}_{\parallel}^{\text{phys}}\right]\!\right] \times \hat{\boldsymbol{n}}, \tag{S3.8b}$$

$$\left[\!\left[D_{\perp}^{\text{phys}}\right]\!\right] = d_{\parallel}\boldsymbol{\nabla}_{\parallel}^{\text{phys}} \cdot \left[\!\left[\boldsymbol{D}_{\parallel}^{\text{phys}}\right]\!\right], \tag{S3.8c}$$

$$\left[\!\left[B_{\perp}^{\text{phys}}\right]\!\right] = 0, \tag{S3.8d}$$

where $[\![\ldots]\!]$ denotes the difference in the electromagnetic fields approaching the interface from both sides. Substituting Eqs. (S3.4)-(S3.7) into Eq. (S3.8), we find

$$\left[\!\left[\boldsymbol{E}_{\parallel}\right]\!\right] = -d_{\perp}\boldsymbol{\nabla}_{\parallel}[\![\Gamma E_{\perp}]\!], \tag{S3.9a}$$

$$\left[\!\left[\boldsymbol{H}_{\parallel}\right]\!\right] = i\omega d_{\parallel}\Gamma\left[\!\left[\boldsymbol{D}_{\parallel}\right]\!\right] \times \hat{\boldsymbol{n}}, \tag{S3.9b}$$

$$[\![D_{\perp}]\!] = d_{\parallel}\Gamma\boldsymbol{\nabla}_{\parallel} \cdot \left[\!\left[\boldsymbol{D}_{\parallel}\right]\!\right], \tag{S3.9c}$$

$$[\![B_{\perp}]\!] = 0. \tag{S3.9d}$$

The above mesoscopic BCs in auxiliary space are applicable to arbitrary systems with isotropic $\boldsymbol{\varepsilon}$ and $\boldsymbol{\mu}$. The proofs of Eq. (S3.9a) and Eq. (S3.9d) are straightforward. We focus on the proofs of the other two BCs. By using Eqs. (S3.4) and (S3.6), Eq. (S3.8b) is rewritten as

$$\begin{aligned}
[\![H_{\parallel,\sigma}]\!] &= i\omega d_{\parallel}\epsilon_{ijk}\Lambda^{i}{}_{\sigma}\Lambda^{j}{}_{\rho}\frac{[\![D_{\parallel}^{\rho}]\!]}{\det\Lambda^{j}{}_{\rho}}g^{km}\hat{n}_{m} \\
&= i\omega d_{\parallel}\epsilon_{\sigma\rho\mu}\Lambda^{\mu}{}_{k}[\![D_{\parallel}^{\rho}]\!]g^{km}\frac{1}{\Gamma}\Lambda^{\nu}{}_{m}\hat{n}_{\nu} \\
&= i\omega d_{\parallel}\epsilon_{\sigma\rho\mu}[\![D_{\parallel}^{\rho}]\!]g^{\mu\nu}\frac{1}{\Gamma}\hat{n}_{\nu}.
\end{aligned} \tag{S3.10}$$

This is Eq. (S3.9b). Similarly, Eq. (S3.8c) is rewritten as

$$\begin{aligned}
[\![D_{\perp}]\!] &= d_{\parallel}\frac{1}{\Gamma^{2}}\Lambda^{\sigma}{}_{i}\partial_{\parallel,\sigma}\left(\Lambda^{i}{}_{\rho}D_{\parallel}^{\rho}\det\Lambda^{\rho}{}_{i}\right) \\
&= d_{\parallel}\Gamma\partial_{\parallel,\sigma}D_{\parallel}^{\rho} + d_{\parallel}\frac{1}{\Gamma^{2}}\left[\Lambda^{\sigma}{}_{i}\left(\partial_{\parallel,\sigma}\Lambda^{i}{}_{\rho}\right)\Gamma^{3}D_{\parallel}^{\rho} + \left(\partial_{\parallel,\rho}\Gamma^{3}\right)D_{\parallel}^{\rho}\right] \\
&= d_{\parallel}\Gamma\partial_{\parallel,i}D_{\parallel}^{i},
\end{aligned} \tag{S3.11}$$

where the two terms within the bracket in the second line cancel each other. This can be proven as follows

$$\begin{aligned}
\partial_{\parallel,\rho}\Gamma^{3} &= \partial_{\rho}\det\Lambda^{\rho}{}_{i} - \hat{n}_{\rho}\left[\frac{g^{\sigma\mu}}{\Gamma^{2}}\hat{n}_{\sigma}\left(\partial_{\mu}\det\Lambda^{\mu}{}_{i}\right)\right] \\
&= -\Gamma^{3}\left\{\partial_{i}\Lambda^{i}{}_{\rho} - \hat{n}_{\rho}\left[\frac{g^{\sigma\mu}}{\Gamma^{2}}\hat{n}_{\sigma}\left(\partial_{i}\Lambda^{i}{}_{\mu}\right)\right]\right\} \\
&= -\Gamma^{3}\left(\partial_{\parallel,i}\Lambda^{i}{}_{\rho}\right).
\end{aligned} \tag{S3.12}$$

In the second line, we have used the Jacobian formula for derivatives of a matrix's determinant with respect to its elements, $\frac{\partial}{\partial M_{ij}}\det\mathbf{M} = (\mathbf{M}^{-1})_{ij}\det\mathbf{M}$, where $\mathbf{M}^{-1}$ is the inverse matrix, and $\left(\partial_{\mu}\Lambda^{\rho}{}_{i}\right)\Lambda^{i}{}_{\rho} = -\left(\partial_{\mu}\Lambda^{i}{}_{\rho}\right)\Lambda^{\rho}{}_{i} = -\left(\partial_{\rho}\Lambda^{i}{}_{\mu}\right)\Lambda^{\rho}{}_{i} = -\partial_{i}\Lambda^{i}{}_{\mu}$. The vector form of Eq. (S3.11) is then nothing but Eq. (S3.9c). So far, we have derived the mesoscopic BCs in the auxiliary space.

### Section IV. Eigenmode expansions and response matrix

At the nanoscale, we adopt the quasi-static approximation, i.e., $E = -\nabla\varphi$. Since $E_{i}^{\mathrm{phys}} = -\partial_{i}\varphi^{\mathrm{phys}}(\boldsymbol{R}) = -\Lambda^{\rho}{}_{i}\partial_{\rho}\varphi(\boldsymbol{r}) = \Lambda^{\rho}{}_{i}E_{\rho}$, the electric potential remains invariant after the transformation, $\varphi^{\mathrm{phys}}(\boldsymbol{R}) = \varphi(\boldsymbol{r})$. Because the dielectric function in the auxiliary space depends on the spatial coordinates, we solve the Poisson equation $\nabla\cdot\varepsilon(\boldsymbol{r})\nabla\varphi(\boldsymbol{r}) = 0$ and employ a mathematical technique, i.e., $\varphi(\boldsymbol{r}) = |\boldsymbol{r}-\boldsymbol{r}_{0}|V(\boldsymbol{r})$ [11], to transform it into a more readily solvable Laplace equation for $V(\boldsymbol{r})$ as

$$\nabla^{2}V(\boldsymbol{r}) = 0. \tag{S4.1}$$

The equation to be solved in physical space is also a Laplace equation $\nabla^2 \varphi^{\mathrm{phys}}(\boldsymbol{R}) = 0$, and thereby, $V(\boldsymbol{r})$ and $\varphi^{\mathrm{phys}}(\boldsymbol{R})$ are inversion potentials of each other [11]. Moreover, we only need to consider Eqs. (S3.9a) and (S3.9c) in auxiliary space, which have been further expressed as

$$[\![\varphi]\!] = -d_{\perp}\Gamma\nabla_{\perp}[\![\varphi]\!], \tag{S4.2a}$$

$$[\![\varepsilon\nabla_{\perp}\varphi]\!] = d_{\parallel}\Gamma\boldsymbol{\nabla}_{\parallel}\cdot[\![\varepsilon\boldsymbol{\nabla}_{\parallel}\varphi]\!]. \tag{S4.2b}$$

The normal basis vector in auxiliary space is $\hat{\boldsymbol{n}} = \hat{\mathbf{e}}_r$ for the interface $\partial\Omega_1$, while $\hat{\boldsymbol{n}} = -\hat{\mathbf{e}}_r$ for the interface $\partial\Omega_2$. Hence, Eq. (S4.2) can be recast as

$$[\![\varphi]\!] = \mp d_{\perp}\Gamma\left[\!\!\left[\frac{\partial}{\partial r}\varphi\right]\!\!\right], \tag{S4.3a}$$

$$\left[\!\!\left[\varepsilon\frac{\partial}{\partial r}\varphi\right]\!\!\right] = \mp d_{\parallel}\left[\!\!\left[\frac{1}{r^2}\frac{\partial}{\partial r}\left(\varepsilon r^2\frac{\partial}{\partial r}\varphi\right)\right]\!\!\right], \tag{S4.3b}$$

in the spherical coordinate system, where $-$ and $+$ denote the interfaces $\partial\Omega_{1,2}$, respectively. By defining the functions $V_1 = \frac{\partial}{\partial r}V$ and $V_2 = \frac{\partial^2}{\partial r^2}V + \frac{2}{r}\frac{\partial}{\partial r}V$, Eq. (S4.3) can be rewritten as

$$[\![V]\!] = \mp\frac{d_{\perp}}{R_T^2}[\![(r - r_0\cos\theta)V + |\boldsymbol{r} - \boldsymbol{r}_0|^2 V_1]\!], \tag{S4.4a}$$

$$\left[\!\!\left[\begin{matrix}\varepsilon^{\mathrm{phys}}(r - r_0\cos\theta)V \\ +\varepsilon^{\mathrm{phys}}|\boldsymbol{r} - \boldsymbol{r}_0|^2 V_1\end{matrix}\right]\!\!\right] = \mp\frac{d_{\parallel}}{R_T^2}\left[\!\!\left[\begin{matrix}\varepsilon^{\mathrm{phys}}\left[3(r_0^2 - r_0^2\cos^2\theta) - \frac{2}{r}r_0\cos\theta\,|\boldsymbol{r} - \boldsymbol{r}_0|^2\right]V \\ +\varepsilon^{\mathrm{phys}}|\boldsymbol{r} - \boldsymbol{r}_0|^4 V_2\end{matrix}\right]\!\!\right], \tag{S4.4b}$$

where $|\boldsymbol{r} - \boldsymbol{r}_0|^2 = r^2 + r_0^2 - 2rr_0\cos\theta$. Equation (S4.4) is the BCs required to match between two materials for Eq. (S4.1).

Regular and irregular solid harmonics are the eigen solutions of Eq. (S4.1) in free space. Hence, $V(\boldsymbol{r})$ in the spherical shell system can be expanded as

$$V(\boldsymbol{r}) = \begin{cases} \sum_{l,m} a_{lm}^{\mathrm{in}}\frac{r^l}{r_0}Y_l^m(\theta,\phi), & r < R_1 \\ \sum_{l,m}\left[(a_{lm}^{+} + a_{lm}^{s+})\frac{r^l}{r_0^l} + a_{lm}^{-}\frac{r_0^{l+1}}{r^{l+1}}\right]Y_l^m(\theta,\phi), & R_0 > r > R_1 \\ \sum_{l,m}\left[a_{lm}^{+}\frac{r^l}{r_0^l} + (a_{lm}^{-} + a_{lm}^{s-})\frac{r_0^{l+1}}{r^{l+1}}\right]Y_l^m(\theta,\phi), & R_2 > r > R_0 \\ \sum_{l,m} a_{lm}^{\mathrm{out}}\frac{r_0^{l+1}}{r^{l+1}}Y_l^m(\theta,\phi), & r > R_2 \end{cases} \tag{S4.5}$$

where $a_{lm}^{\text{in/out}}$, $a_{lm}^{\pm}$, and $a_{lm}^{s\pm}$ ($l=0,1,2,\cdots$ and $m=-l,\cdots,0,\cdots,+l$) are the expansion coefficients, as indicated in Fig. 1(a), and thereby

$$V_1 = \begin{cases} \sum_{l,m} a_{lm}^{\text{in}} \frac{l}{r}\frac{r^l}{R_0^l} Y_l^m(\theta,\phi), & r < R_1 \\ \sum_{l,m} \left[ (a_{lm}^{+} + a_{lm}^{s+}) \frac{l}{r}\frac{r^l}{R_0^l} - a_{lm}^{-} \frac{l+1}{r}\frac{R_0^{l+1}}{r^{l+1}} \right] Y_l^m(\theta,\phi), & R_0 > r > R_1 \\ \sum_{l,m} \left[ a_{lm}^{+} \frac{l}{r}\frac{r^l}{R_0^l} - (a_{lm}^{-} + a_{lm}^{s-}) \frac{l+1}{r}\frac{R_0^{l+1}}{r^{l+1}} \right] Y_l^m(\theta,\phi), & R_2 > r > R_0 \\ \sum_{l,m} a_{lm}^{\text{out}} \left(-\frac{l+1}{r}\right) \frac{R_0^{l+1}}{r^{l+1}} Y_l^m(\theta,\phi), & r > R_2 \end{cases} \tag{S4.6}$$

$$V_2 = \begin{cases} \sum_{l,m} a_{lm}^{\text{in}} \frac{l(l+1)}{r^2}\frac{r^l}{R_0^l} Y_l^m(\theta,\phi), & r < R_1 \\ \sum_{l,m} \left[ (a_{lm}^{+} + a_{lm}^{s+}) \frac{l(l+1)}{r^2}\frac{r^l}{R_0^l} + a_{lm}^{-} \frac{l(l+1)}{r^2}\frac{R_0^{l+1}}{r^{l+1}} \right] Y_l^m(\theta,\phi), & R_0 > r > R_1 \\ \sum_{l,m} \left[ a_{lm}^{+} \frac{l(l+1)}{r^2}\frac{r^l}{R_0^l} + (a_{lm}^{-} + a_{lm}^{s-}) \frac{l(l+1)}{r^2}\frac{R_0^{l+1}}{r^{l+1}} \right] Y_l^m(\theta,\phi), & R_2 > r > R_0 \\ \sum_{l,m} a_{lm}^{\text{out}} \frac{l(l+1)}{r^2} \frac{R_0^{l+1}}{r^{l+1}} Y_l^m(\theta,\phi), & r > R_2 \end{cases} \tag{S4.7}$$

Equations (S4.4a)-(S4.4b) contain $\cos\theta$ terms up to the second order, and the relevant recursive relations for the spherical harmonics associated with $\cos\theta$ are

$$\cos\theta\, Y_l^m = p_{l+1}^m Y_{l+1}^m + p_l^m Y_{l-1}^m, \tag{S4.8a}$$

$$\cos^2\theta\, Y_l^m = p_{l+2}^m p_{l+1}^m Y_{l+2}^m + [(p_{l+1}^m)^2 + (p_l^m)^2] Y_l^m + p_l^m p_{l-1}^m Y_{l-2}^m, \tag{S4.8b}$$

where $p_l^m = \sqrt{\frac{l^2-m^2}{4l^2-1}}$. Now, we match the BCs Eqs. (S4.4a)-(S4.4b). Observing Eqs. (S4.4) and (S4.8), we find that the azimuthal quantum number $m$ is conserved, and thereby, substituting Eqs. (S4.5)-(S4.8) into Eq. (S4.4) leads to

$$\begin{aligned} &\sum_{l'} \left( a_{l'm}^{\text{in/out}} - a_{l'm}^{\pm} - a_{l'm}^{s\pm} \right) \left[ \delta_{l,l'} - \frac{d_\perp}{\chi_0} S_{l,l'}^{\pm\pm} \right] \\ &= \sum_{l'} a_{l'm}^{\mp} \chi_\pm^{-(2l'+1)} \left[ \delta_{l,l'} + \frac{d_\perp}{\chi_0} \left[ S_{l,l'}^{\pm\pm} - \left( \chi_\pm^{-1} - \chi_\pm \right) \delta_{l,l'} \right] \right], \end{aligned} \tag{S4.9a}$$

$$\begin{aligned} &\sum_{l'} \left[ \varepsilon_{\text{m}} a_{l'm}^{\text{in/out}} - \varepsilon_{\text{d}} \left( a_{l'm}^{\pm} + a_{l'm}^{s\pm} \right) \right] \left[ S_{l,l'}^{\pm\pm} - \frac{d_\parallel}{\chi_0} V_{l,l'}^{\pm\pm} \right] \\ &= -\sum_{l'} \varepsilon_{\text{d}} a_{l'm}^{\mp} \chi_\pm^{-(2l'+1)} \left[ \left[ S_{l,l'}^{\pm\pm} - \left( \chi_\pm^{-1} - \chi_\pm \right) \delta_{l,l'} \right] + \frac{d_\parallel}{\chi_0} V_{l,l'}^{\pm\pm} \right], \end{aligned} \tag{S4.9b}$$

where

$$S_{l,l'}^{\pm\pm} = -\left[(l'+1)\chi_{\pm} + l'\chi_{\pm}^{-1}\right]\delta_{l,l'} + (2l'+1)\left(p_{l'+1}\chi_{\pm}^{-1}\delta_{l,l'+1} + p_{l'}\chi_{\pm}\delta_{l,l'-1}\right), \tag{S4.10}$$

$$\begin{aligned} V_{l,l'}^{\pm\pm} = &\left[3 + [4l'(l'+1)+1]\left[\left(p_{l'+1}^{m}\right)^2 + \left(p_{l'}^{m}\right)^2\right] + l'(l'+1)\left(\chi_{\pm} + \chi_{\pm}^{-1}\right)^2\right]\delta_{l,l'} \\ &-[4l'(l'+1)+2]\left(\chi_{\pm} + \chi_{\pm}^{-1}\right)\left(p_{l'+1}^{m}\chi_{\pm}^{-1}\delta_{l,l'+1} + p_{l'}^{m}\chi_{\pm}\delta_{l,l'-1}\right) \\ &+[4l'(l'+1)+1]\left(p_{l'+2}^{m}p_{l'+1}^{m}\chi_{\pm}^{-2}\delta_{l,l'+2} + p_{l'}^{m}p_{l'-1}^{m}\chi_{\pm}^{2}\delta_{l,l'-2}\right). \end{aligned} \tag{S4.11}$$

We write the above equations in terms of matrices

$$\left(\mathbf{I} - \frac{d_{\perp}}{L_0}\mathbf{S}_{m}^{\pm\pm}\right)\left[\boldsymbol{a}_{m}^{\text{in/out}} - \left(\boldsymbol{a}_{m}^{\pm} + \boldsymbol{a}_{m}^{s\pm}\right)\right] = \left[\mathbf{I} + \frac{d_{\perp}}{L_0}\mathbf{W}_{m}^{\pm}\right]\mathbf{D}^{\pm}\boldsymbol{a}_{m}^{\mp}, \tag{S4.12a}$$

$$\left(\mathbf{S}_{m}^{\pm\pm} - \frac{d_{\parallel}}{L_0}\mathbf{V}_{m}^{\pm\pm}\right)\left[\varepsilon_{\text{m}}\boldsymbol{a}_{m}^{\text{in/out}} - \varepsilon_{\text{d}}\left(\boldsymbol{a}_{m}^{\pm} + \boldsymbol{a}_{m}^{s\pm}\right)\right] = -\varepsilon_{\text{d}}\left[\mathbf{W}_{m}^{\pm} + \frac{d_{\parallel}}{L_0}\mathbf{V}_{m}^{\pm\pm}\right]\mathbf{D}^{\pm}\boldsymbol{a}_{m}^{\mp}, \tag{S4.12b}$$

where $\mathbf{W}_{m}^{\pm} = \mathbf{S}_{m}^{\pm\pm} - \left(\chi_{\pm}^{-1} - \chi_{\pm}\right)\mathbf{I}$ and $D_{l,l'}^{\pm} = \chi_{\pm}^{-(2l'+1)}\delta_{l,l'}$. Combining Eqs. (S4.12a)-(S4.12b), we obtain the response matrix equation

$$\begin{pmatrix} -\mathbf{I} & {\mathbf{S}_{m,\text{wd}}^{++}}^{-1}\mathbf{T}_{m,\text{wd}}^{+-} \\ {\mathbf{S}_{m,\text{wd}}^{--}}^{-1}\mathbf{T}_{m,\text{wd}}^{-+} & -\mathbf{I} \end{pmatrix}\begin{pmatrix} \boldsymbol{a}_{m}^{+} \\ \boldsymbol{a}_{m}^{-} \end{pmatrix} = \begin{pmatrix} \boldsymbol{a}_{m}^{s+} \\ \boldsymbol{a}_{m}^{s-} \end{pmatrix}, \tag{S4.13}$$

where

$$\mathbf{S}_{m,\text{wd}}^{\pm\pm} = \mathbf{S}_{m}^{\pm\pm}\left(\mathbf{I} - \frac{d_{\perp}}{L_0}\mathbf{S}_{m}^{\pm\pm} - \frac{d_{\parallel}}{L_0}{\mathbf{S}_{m}^{\pm\pm}}^{-1}\mathbf{V}_{m}^{\pm\pm}\right), \tag{S4.14}$$

$$\mathbf{T}_{m,\text{wd}}^{\pm\mp} = \mathbf{T}_{m}^{\pm\mp} + \frac{d_{\perp}}{L_0}\mathbf{S}_{m}^{\pm\pm}\left[\left(\chi_{\pm}^{-1} - \chi_{\pm}\right)\mathbf{I} - \mathbf{S}_{m}^{\pm\pm}\right]\mathbf{D}^{\pm} + \frac{d_{\parallel}}{L_0}\mathbf{V}_{m}^{\pm\pm}\mathbf{D}^{\pm}, \tag{S4.15}$$

$$\mathbf{T}_{m}^{\pm\mp} = \left[\frac{\varepsilon_{\text{d}}}{(\varepsilon_{\text{m}} - \varepsilon_{\text{d}})}\left(\chi_{\pm}^{-1} - \chi_{\pm}\right)\mathbf{I} - e^{-\alpha}\mathbf{S}_{m}^{\pm\pm}\right]\mathbf{D}^{\pm}. \tag{S4.16}$$

The matrices $\mathbf{T}_{m,\text{wd}}$ and $\mathbf{S}_{m,\text{wd}}$ with the "wd" subscript contain additional matrix terms modified by *d*-parameters. When $d_{\perp,\parallel} = 0$, they revert to $\mathbf{T}_{m}$ and $\mathbf{S}_{m}$ classically [3]. The 2×2 block matrix in Eq. (S4.13) is a response matrix, which is determined by geometric parameters, bulk material properties, and surface response functions $d_{\sigma}$. Notably, ${\mathbf{S}_{m,\text{wd}}^{++}}^{-1}\mathbf{T}_{m,\text{wd}}^{+-}$ and ${\mathbf{S}_{m,\text{wd}}^{--}}^{-1}\mathbf{T}_{m,\text{wd}}^{-+}$ are the response matrix of $\Omega_1$ and $\Omega_2$ individually. $(\boldsymbol{a}_{m}^{s+} \quad \boldsymbol{a}_{m}^{s-})^{\text{T}}$ is the column vector formed by the coefficients of the excitation source, while $(\boldsymbol{a}_{m}^{+} \quad \boldsymbol{a}_{m}^{-})^{\text{T}}$ represents the scattering fields. In principle, the dimensions of these matrices are infinite. However, in practical numerical computations, they are truncated at $l = l_c$. Thus, the matrix dimension for each $m$ in Eqs. (S4.14)-(S4.16) is $(l_c - |m| + 1) \times (l_c - |m| + 1)$.

## Section V. Absorption spectrum and mode profile

Aiming to validate the 3D-CM method, we first present the fitted expressions of the Feibelman *d*-parameters for Na and Ag in Sec. V A and consistent bulk parameters in Sec. V B. Subsequently, in Sec. V C, how to calculate the absorption spectrum and mode profile by our 3D-CM method is given and examined with finite element method (FEM). At last, in Sec. V D, the 3D-CM method with radiation correction is presented and works well at the nanoscale scale of interest.

### A. Expressions of Feibelman *d*-parameters

The parameters $d_\perp$ and $d_\parallel$, respectively, characterizing the centroids of the induced electron density and the normal derivative of the current density, shall be obtained through time-dependent density functional theory or self-consistent hydrodynamic model (SCHDM). Due to the causality of QSRs automatically required to satisfy the Kramers-Kronig relation, the $d_\perp$ data is fitted by the modified Lorentz resonance terms as

$$d_\perp(\omega) = \sum_j \frac{f_j - i\gamma_j'\omega}{\omega_j^2 - \omega(\omega + i\gamma_j)}, \tag{S5.1}$$

where the unit of $\omega_j$ and $\gamma_j$ ($f_j$ and $\gamma_j'$) is eV ($\mathrm{nm \cdot eV^2}$).

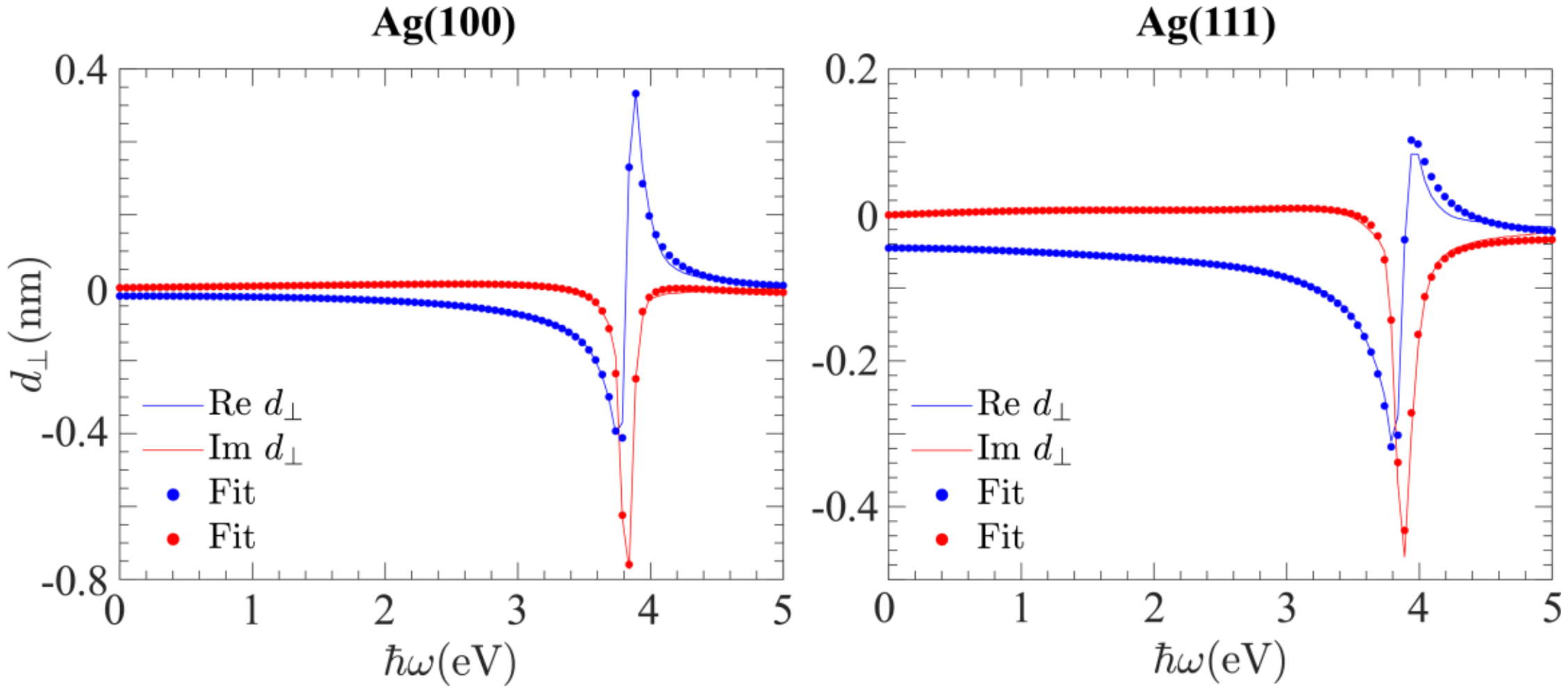

**Figure S3**. $d_\perp(\omega)$ from calculation data in Ref. [12] (solid lines) and fitting formula (filled circles). The left and right panels, respectively, show the Ag(100) and Ag(111) cases.

**Table S1**. Fitting parameters in Eq. (S5.1) for Ag(100) and Ag(111).

| Ag(100) | | | | |
|---|---|---|---|---|
| $j$ | $f_j$ | $\gamma_j'$ | $\omega_j$ | $\gamma_j$ |
| 1 | -0.02485 | -0.0883 | 4.148645 | 2.594482 |
| 2 | -0.00606 | 0.001514 | 3.884249 | 0.013458 |
| 3 | -0.30622 | 0.008441 | 3.821321 | 0.089689 |
| **Ag(111)** | | | | |
| $j$ | $f_j$ | $\gamma_j'$ | $\omega_j$ | $\gamma_j$ |
| 1 | -0.57189 | -0.35423 | 4.130138 | 8.22803 |
| 2 | 0.043227 | -0.01596 | 3.514258 | 1.276212 |
| 3 | -0.22827 | -0.01438 | 3.867799 | 0.137907 |

For a typical alkali metal Na, $d_\perp$ from SCHDM can be fitted by a single term, with the fitting parameters $f_1 = 1.13569$ nm·eV$^2$, $\gamma_1' = 0$ nm·eV$^2$, $\omega_1 = 4.55$ eV, and $\gamma_1 = 0.177$ eV [13]. $d_\perp$ of Ag(100) and Ag(111) calculated in Ref. [12] are nicely fitted by Eq. (S5.1) with three terms, as exhibited in Figure S3. The corresponding fitting parameters are presented in Table S1.

**B. Expressions of permittivity consistent with *d*-parameters**

The bulk permittivity shall be consistent with the employed *d*-parameters [12,13]. For Na, we use a Drude model $\epsilon_m = 1 - \frac{\omega_p^2}{\omega(\omega+i\gamma)}$ with $\hbar\omega_p = 5.89$ eV and $\hbar\gamma = 0.17$ eV, as in Ref. [13]. For Ag, according to Ref. [12], we first obtain a background permittivity $\epsilon_b = \epsilon_m^{\mathrm{exp}} + \frac{\left(\omega_p^{\mathrm{exp}}\right)^2}{\omega(\omega+i\gamma^{\mathrm{exp}})}$, where $\epsilon_m^{\mathrm{exp}}$ is the experimental tabulated data, $\omega_p^{\mathrm{exp}} = 9.17$ eV, and $\gamma^{\mathrm{exp}} = 0.21$ eV. Here, $\epsilon_m^{\mathrm{exp}}$ is taken from Ref. [14] for the dominant frequency range 0.64-6.6 eV, as in Ref. [12], and otherwise from Ref. [15] due to the integral of the Lifshitz formula involving the ultrahigh frequency range. For distinct crystal faces, $\epsilon_b$ is then $\epsilon_m = \epsilon_b - \frac{\left(\omega_p^{\mathrm{ALP}}\right)^2}{\omega(\omega+i\gamma^{\mathrm{exp}})}$ where $\hbar\omega_p^{\mathrm{ALP}} = 8.80$ eV for the (100) facet and $\hbar\omega_p^{\mathrm{ALP}} = 9.19$ eV for the (111) facet. To obtain an analytical expression for permittivity $\epsilon_m$, we fit $\epsilon_b$ using the modified Lorentz resonance terms as

$$\epsilon_b(\omega) = 1 - \sum_j \frac{f_j - i\gamma_j'\omega}{\omega_j^2 - \omega\left(\omega + i\gamma_j\right)}. \tag{S5.2}$$

Figure S4 shows the fitting results of Ag, and the corresponding parameters are provided in Table S2.

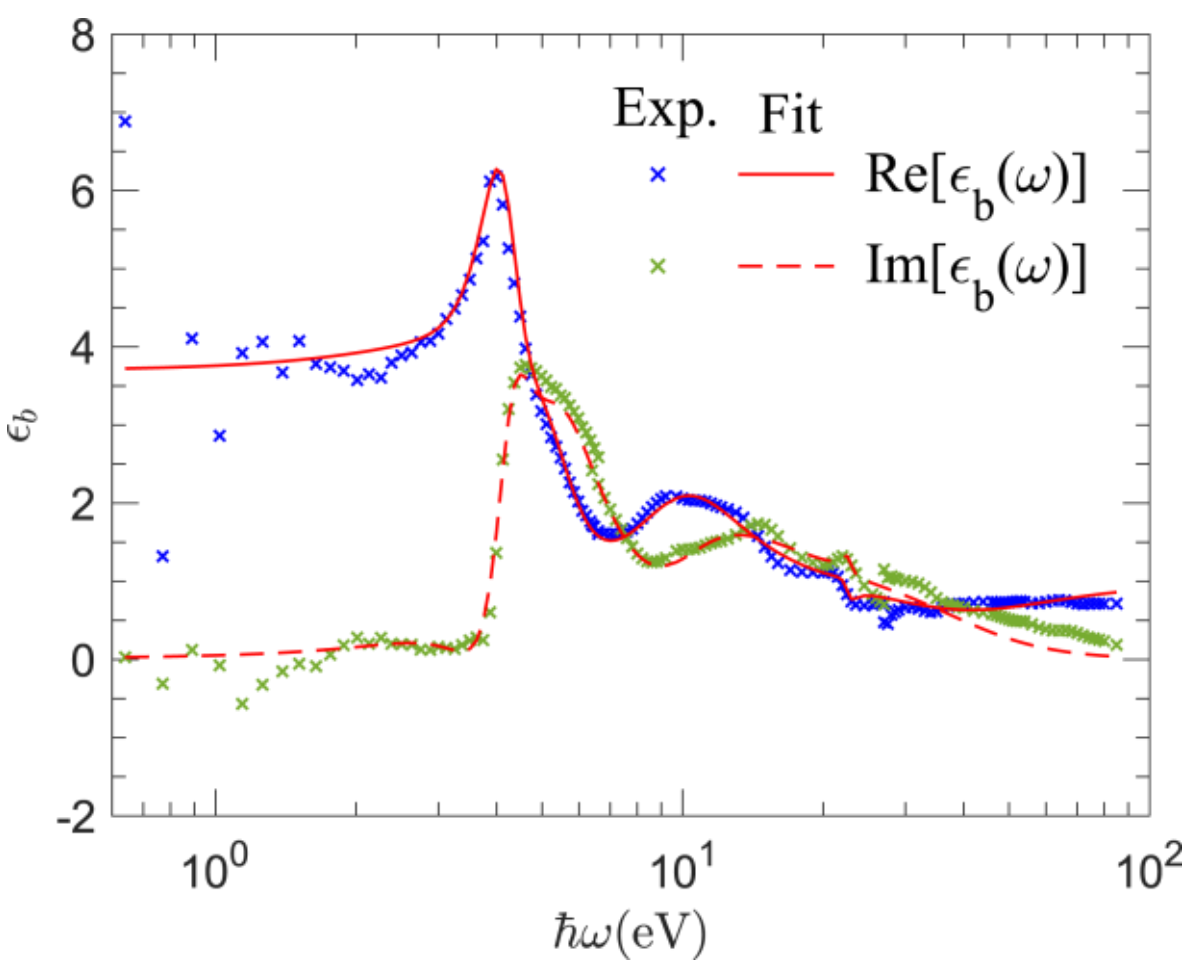

**Figure S4.** Background permittivity $\epsilon_b$ of Ag as a function of frequency $\omega$. The crosses show the experimental data, while the solid and dashed lines show the results from the fitting formula.

**Table S2**. Fitting parameters in Eq. (S5.2) for Ag.

| Ag | | | | |
|---|---|---|---|---|
| $j$ | $f_j$ | $\gamma_j'$ | $\omega_j$ | $\gamma_j$ |
| 1 | -19.9864 | -12.0827 | 2.603252 | 5.071371 |
| 2 | -15.4395 | -4.83264 | 1.184436 | 4.20529 |
| 3 | -15.7406 | 10.58498 | 3.088868 | 4.22522 |
| 4 | -14.7739 | -12.8479 | 8.444382 | 10.22816 |
| 5 | -5.91085 | 0.174586 | 1.400384 | 22.67762 |
| 6 | -19.0993 | 19.70987 | 49.99987 | 39.93956 |

**C. Absorption spectrum and mode profile**

With the fitted $d$-parameters and consistent bulk parameters, we now investigate absorption spectra and mode profiles using our 3D-CM method. The plane wave $\boldsymbol{E}_0^{\text{phys}}$ in physical space is transformed into an electric dipole

$$\boldsymbol{p} = 4\pi\varepsilon_0 R_T^3 \boldsymbol{E}_0^{\text{phys}}, \tag{S5.3}$$

at the inversion point in the auxiliary space [1], as illustrated in Fig. S2. $\boldsymbol{E}_0^{\text{phys}}$ and $\boldsymbol{p}$ oriented to arbitrary direction can be decomposed into components parallel and perpendicular to the $z$-axis. Next, we will give derivations of the absorption spectra for such two cases separately.

If $\boldsymbol{p} = p\hat{\boldsymbol{e}}_z$ is parallel to the $z$-axis, the electric potential can be expanded in terms of spherical harmonics as

$$
\begin{aligned}
V_{\boldsymbol{p}} &= \frac{p}{4\pi\varepsilon_0}\frac{R_0 - r\cos\theta}{|\boldsymbol{r}-\boldsymbol{R}_0|^3} \\
&= \frac{p}{4\pi\varepsilon_0 R_0^2}\begin{cases} \displaystyle\sum_{l=0}^{\infty}(l+1)\sqrt{\frac{4\pi}{2l+1}}\left(\frac{r}{R_0}\right)^{l} Y_l^0(\cos\theta) & r < R_0, \\ \displaystyle\sum_{l=0}^{\infty}(-l)\sqrt{\frac{4\pi}{2l+1}}\left(\frac{R_0}{r}\right)^{l+1} Y_l^0(\cos\theta) & r > R_0. \end{cases}
\end{aligned} \tag{S5.4}
$$

The expansion coefficients of source electric potential $\boldsymbol{a}_s^{\pm}$ are then obtained by comparing with Eq. (S4.5). The expansion coefficients of the induced electric potential $\boldsymbol{a}^{\pm}$ are straightforwardly obtained by using Eq. (S4.13), and the scattering field at the dipole $\boldsymbol{p}$ in the auxiliary space is

$$
\boldsymbol{E}_{\mathrm{sca}}(\boldsymbol{R}_0) = \nabla V_{\mathrm{sca}}|_{r=R_0,\theta=0}. \tag{S5.5}
$$

For simplicity, we define the reduced expansion coefficients as

$$
\tilde{\boldsymbol{a}}_s^{\pm} = \frac{4\pi\varepsilon_0 R_0^2}{p}\boldsymbol{a}_s^{\pm}, \qquad \tilde{\boldsymbol{a}}^{\pm} = \frac{4\pi\varepsilon_0 R_0^2}{p}\boldsymbol{a}^{\pm}. \tag{S5.6}
$$

The power of absorption is the work per second done by the scattered fields $\boldsymbol{E}_{\mathrm{sca}}$ on $\boldsymbol{p}$ as

$$
\begin{aligned}
P_{\mathrm{abs}} &= \frac{\omega}{2}\mathrm{Im}[\boldsymbol{p}^* \cdot \boldsymbol{E}_{\mathrm{sca}}(R_0, 0)] \\
&= \frac{|p|^2\omega}{8\pi\varepsilon_0 R_0^2}\mathrm{Im}\sum_{l=0}^{\infty}\left(\frac{l}{R_0}\tilde{a}_{lm}^{+} - \frac{l+1}{R_0}\tilde{a}_{lm}^{-}\right)Y_l^0(0,\phi) \\
&= 2\pi\varepsilon_0 L_0^3\left|\boldsymbol{E}_0^{\mathrm{phys}}\right|^2\omega\,\mathrm{Im}\sum_{l=0}^{\infty}(l\tilde{a}_{lm}^{+} - (l+1)\tilde{a}_{lm}^{-})Y_l^0(0,\phi)\,.
\end{aligned} \tag{S5.7}
$$

Finally, the absorption cross-section is

$$
\sigma_{\mathrm{abs}} = \frac{P_{\mathrm{abs}}}{I_0} = 4\pi L_0^3 k_0\,\mathrm{Im}\sum_{l=0}^{\infty}[l\tilde{a}_{lm}^{+} - (l+1)\tilde{a}_{lm}^{-}]Y_l^0(0,\phi) \tag{S5.8}
$$

where $I_0 = \frac{1}{2}\varepsilon_0 c\left|\boldsymbol{E}_0^{\mathrm{phys}}\right|^2$ is the incident power flux and $k_0 = \omega/c$ is the wave number.

Now, we turn to the case that $\boldsymbol{p}$ is perpendicular to the $z$-axis. Due to the symmetry of rotation along the $z$-axis, we assume $\boldsymbol{p} = p\hat{\boldsymbol{e}}_x$ is oriented to the $x$-axis. The electric potential can be expanded in terms of spherical harmonics as

$$V_{\boldsymbol{p}}=\frac{p}{4\pi\varepsilon_0}\frac{r\sin\theta\cos\phi}{|\boldsymbol{r}-\boldsymbol{R}_0|^3}=-\frac{p\cos\phi}{4\pi\varepsilon_0 R_0}\begin{cases}\sum_{l=0}^{\infty}\frac{r^l}{R_0^{l+1}}\frac{\partial}{\partial\theta}P_l(\cos\theta),r<R_0,\\ \sum_{l=0}^{\infty}\frac{R_0^l}{r^{l+1}}\frac{\partial}{\partial\theta}P_l(\cos\theta),r>R_0.\end{cases}\tag{S5.9}$$

Employing a useful property of the associated Legendre polynomials

$$\begin{aligned}\frac{\partial}{\partial\theta}P_l(\cos\theta)&=-\sin\theta\frac{\partial}{\partial\cos\theta}P_l(\cos\theta)\\ &=-\frac{1}{2}[-P_l^1(\cos\theta)+l(l+1)P_l^{-1}(\cos\theta)],\end{aligned}\tag{S5.10}$$

and an identity $Y_l^m(\theta,\phi)=\sqrt{\frac{2l+1}{4\pi}\frac{(l-m)!}{(l+m)!}}P_l^m(\cos\theta)e^{im\phi}$, we find

$$\frac{\partial}{\partial\theta}P_l(\cos\theta)=-\frac{1}{2}\sqrt{\frac{4\pi}{2l+1}}(l+1)l\left(Y_l^1(\theta,\phi)-Y_l^{-1}(\theta,\phi)\right),\tag{S5.11}$$

and subsequently

$$V_{\boldsymbol{p}}=-\frac{p}{4\pi\varepsilon_0 R_0^2}\begin{cases}\sum_{l=0}^{\infty}\frac{r^l}{R_0^l}\frac{1}{2}\sqrt{\frac{4\pi}{2l+1}}(l+1)l\left(Y_l^1(\theta,\phi)-Y_l^{-1}(\theta,\phi)\right),r<R_0,\\ \sum_{l=0}^{\infty}\frac{R_0^{l+1}}{r^{l+1}}\frac{1}{2}\sqrt{\frac{4\pi}{2l+1}}(l+1)l\left(Y_l^1(\theta,\phi)-Y_l^{-1}(\theta,\phi)\right),r>R_0.\end{cases}\tag{S5.12}$$

Similar to Eqs. (S5.7)-(S5.8), the absorption power is

$$P_{\text{abs}}=2\pi\varepsilon_0 L_0^3\left|\boldsymbol{E}_0^{\text{phys}}\right|^2\omega\,\text{Im}\sum_{l=1}^{\infty}(\tilde{a}_{l\,1}^{+}+\tilde{a}_{l\,1}^{-})\sqrt{\frac{2l+1}{4\pi}}(l+1)l,\tag{S5.13}$$

and the absorption cross-section is

$$\sigma_{\text{abs}}=4\pi L_0^3 k_0\,\text{Im}\sum_{l=0}^{\infty}(\tilde{a}_{l\,1}^{+}+\tilde{a}_{l\,1}^{-})\sqrt{\frac{2l+1}{4\pi}}(l+1)l.\tag{S5.14}$$

We utilize the sodium dimer system as a prototype to showcase the absorption spectra [Eqs. (S5.8) and (S5.14)] and mode profiles [Eq. (S5.5)] using our 3D-CM method. The geometric parameters are $R_1=R_2=5$ nm and $\delta=1$ nm. $d_\perp$ is displayed in Fig. 2(a), and $d_\parallel=0$ due to the neutrality. Figures S5(a)-(c) and S5(d)-(f) show the results for two orthogonal polarizations, with the direction of the incident electric field shown in the insets of Figs. S5(a) and S5(d). By comparing the 3D-CM method with FEM developed in Ref. [4], we find that our analytical approach almost perfectly matches the FEM results in terms of both the absorption cross-section and the scattered field distributions, as well as finer details of the scattered field. The absorption cross-section for arbitrary polarizations is a linear superposition of two orthogonal polarizations, which is thus obtainable as well. Moreover, the matrix dimension is always smaller than 100 in Fig. S5 with total degrees of freedom (DOFs) less than 10k, but the DOF of FEM requires about 5M to have the convergent results in Fig. S5. All of these highlight the accuracy and efficiency of our 3D-CM method.

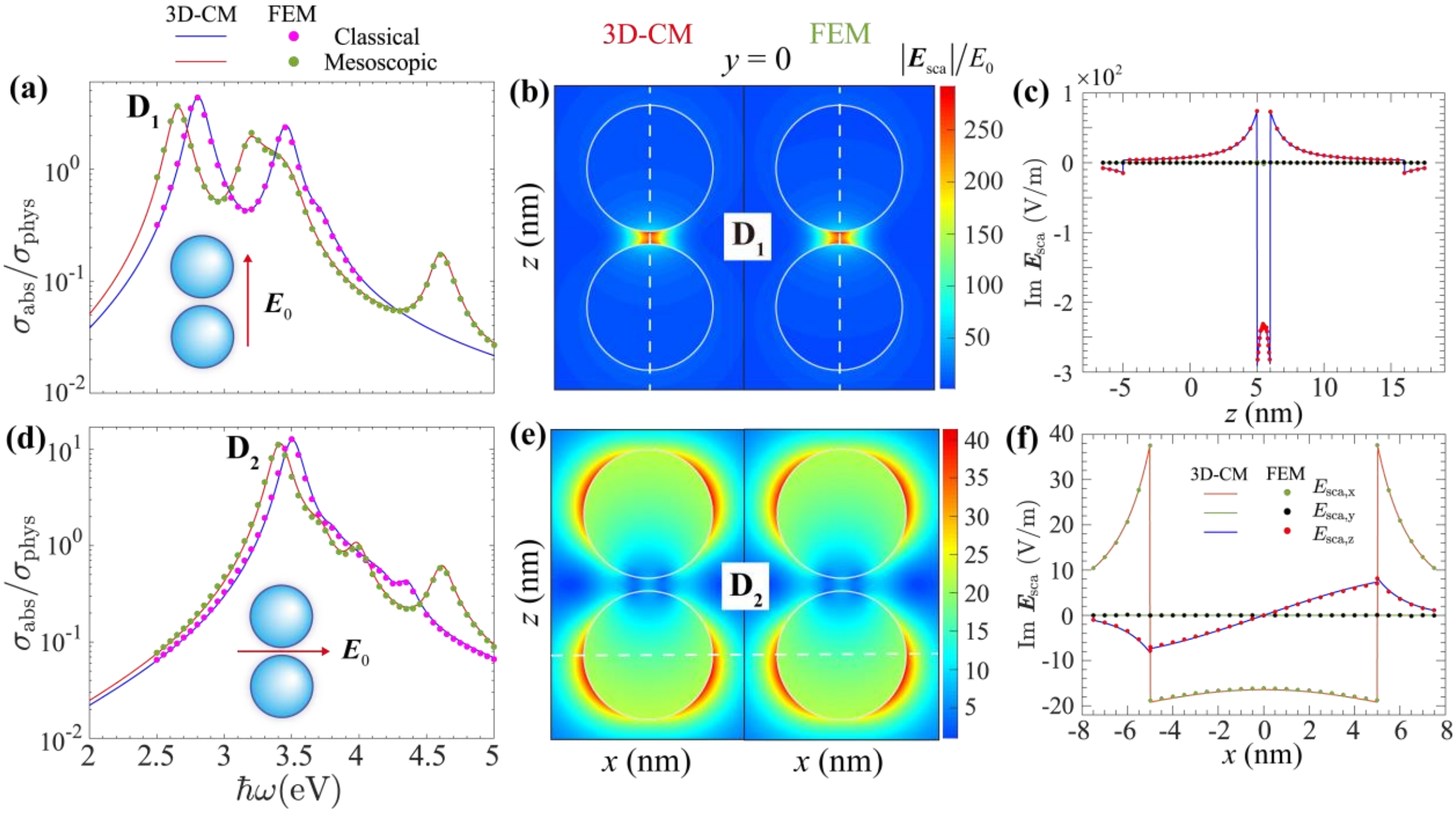

**Figure S5**. (a, d) Absorption cross-section $\sigma_{\text{abs}}$ spectra (normalized by the physical cross-section $\sigma_{\text{phys}}$) for two perpendicular polarizations indicated by the red arrow ($\boldsymbol{E}_0$) in the insets. $\sigma_{\text{phys}} = 2\pi R_1^2$ in (a), while $\sigma_{\text{phys}} = \pi R_1^2$ in (d). The $D_1$ and $D_2$ mark the lowest-frequency resonance peaks in each polarization. (b, e) The magnitude of the scattered fields $|\boldsymbol{E}_{\text{sca}}|$ in the $y = 0$ plane at $D_1$ and $D_2$, normalized by the incident field magnitude $E_0$, is depicted by the color plot. The left and right panels correspond to the 3D-CM method and FEM, respectively. (c, f) $\text{Im}(\boldsymbol{E}_{\text{sca}})$ along the white dashed lines in (b) and (e). Solid lines (filled circles) in (a,c,d,f) represents the 3D-CM method (FEM). The truncation order $l_c$ is 35 here.

### D. Radiation correction

When the size of nanoparticles is small, Eqs. (S5.8) and (S5.14) are accurate. However, when the size is larger, the retardation effect (radiation loss) shall be included. To embed the retardation effect in the 3D-CM, we have included the dipole radiation corrections in the effective polarizability. Specifically, since $\boldsymbol{p}$ can be expressed as the product of a polarizability $\alpha$ and $\boldsymbol{E}_{\text{sca}}(\boldsymbol{R}_0)$ in auxiliary space, i.e.,

$$\boldsymbol{p} = \alpha \boldsymbol{E}_{\text{sca}}(\boldsymbol{R}_0), \tag{S5.15}$$

we correct $\alpha$ to be an effective polarizability $\alpha_{\text{eff}}$. The following are detailed derivations.

Of interest is that $\boldsymbol{E}_{\text{sca}}(\boldsymbol{R}_0)$ can be linked to a dipole $\boldsymbol{p}_{\text{s}}^{\text{phys}}$ at the origin in physical space, just as $\boldsymbol{p}$ relates to $\boldsymbol{E}_0^{\text{phys}}$ by Eq. (S5.3), i.e.,

$$\boldsymbol{E}_{\text{sca}}(\boldsymbol{R}_0) = 4\pi\varepsilon_0 R_T^3 \boldsymbol{p}_{\text{s}}^{\text{phys}}. \quad \text{(S5.16)}$$

On other aspects, $\boldsymbol{p}_{\text{s}}^{\text{phys}}$ relates to a scattered field $\boldsymbol{E}_{\text{sca}}^{\text{phys}}$ by a dyadic Green tensor $\boldsymbol{G}$, i.e.,

$$\boldsymbol{E}_{\text{s}}^{\text{phys}}(\boldsymbol{r}) = i\omega^2\mu_0 \boldsymbol{G}(\boldsymbol{r}, \boldsymbol{r}_0) \cdot \boldsymbol{p}_{\text{s}}^{\text{phys}}, \quad \text{(S5.17)}$$

where the dyadic Green tensor is expressed as

$$\boldsymbol{G}(\boldsymbol{r}, \boldsymbol{r}_0) = \left(I + \frac{1}{k_0^2}\nabla\nabla\right)\frac{e^{ik_0|\boldsymbol{r}-\boldsymbol{r}_0|}}{4\pi|\boldsymbol{r}-\boldsymbol{r}_0|}. \quad \text{(S5.18)}$$

Only the imaginary part of $\boldsymbol{G}$ contributes to the radiative damping. Under the near-field approximation ($k_0|\boldsymbol{r}-\boldsymbol{r}_0| \ll 1$),

$$\operatorname{Im}\boldsymbol{G}(\boldsymbol{r} \to \boldsymbol{r}_0, \boldsymbol{r}_0) = \frac{k_0}{6\pi}\boldsymbol{I}. \quad \text{(S5.19)}$$

Hence, we obtain

$$\boldsymbol{E}_{\text{s}}^{\text{phys}}(\boldsymbol{r}) = i\frac{k_0^3}{6\pi\varepsilon_0}\boldsymbol{p}_{\text{s}}^{\text{phys}}. \quad \text{(S5.20)}$$

It can be seen that a dipole radiates a uniform scattered near-field. Moreover, the uniform field $\boldsymbol{E}_{\text{s}}^{\text{phys}}$ can relate to the dipole $\boldsymbol{p}_{\text{s}}$ of a fictive absorber in auxiliary space

$$\boldsymbol{p}_{\text{s}} = 4\pi\varepsilon_0 R_T^3 \boldsymbol{E}_{\text{s}}^{\text{phys}} = i\frac{2k_0^3 R_T^3}{3}\boldsymbol{p}_{\text{s}}^{\text{phys}}. \quad \text{(S5.21)}$$

Inserting Eq. (S5.21) into Eq. (S5.20), we can obtain

$$\boldsymbol{p}_{\text{s}} = \gamma_{\text{s}}\boldsymbol{E}_{\text{sca}}(\boldsymbol{R}_0), \quad \text{(S5.22)}$$

where $\gamma_{\text{s}} = i\frac{8}{3}\pi\epsilon_0 k_0^3 R_T^6$. Hence, the radiative loss can be modeled as the absorbed power from the absorber of a polarizability $\gamma_{\text{s}}$ in auxiliary space. By adding the contribution of the absorber, i.e., $\boldsymbol{p} + \boldsymbol{p}_{\text{s}} = \alpha\boldsymbol{E}_{\text{sca}}(\boldsymbol{R}_0)$, Eq. (S5.15) is rewritten as

$$\boldsymbol{p} = \alpha_{\text{eff}}\boldsymbol{E}_{\text{sca}}(\boldsymbol{R}_0), \quad \text{(S5.23)}$$

where $\alpha_{\text{eff}} = \alpha - \gamma_{\text{s}}$ is the effective polarizability. Finally, by inserting Eq. (S5.23) into Eqs. (S5.7)-(S5.8) and (S5.13)-(S5.14), we can obtain the absorption cross-section with radiation correction.

Figure S6(a) shows the radiation-corrected (RC) absorption spectra $\sigma_{\text{abs}}$ (color contour) for various sphere radii $R$. Figures S6(b)-(c) compare the absorption spectra at three fixed radii with the FEM results. We see that when $R = 5$ nm and 10 nm, the RC 3D-CM method works nicely but will have some deviations when $R$ goes to 20 nm. The first two orders of LSP resonances and Bennett modes calculated using FEM are labeled by open circles in Fig. S6(a). This indicates that the LSPs are indeed impacted by the retardation effect for large $R$, but our RC 3D-CM method works in the nanoscale of interest ($L < 10$ nm).

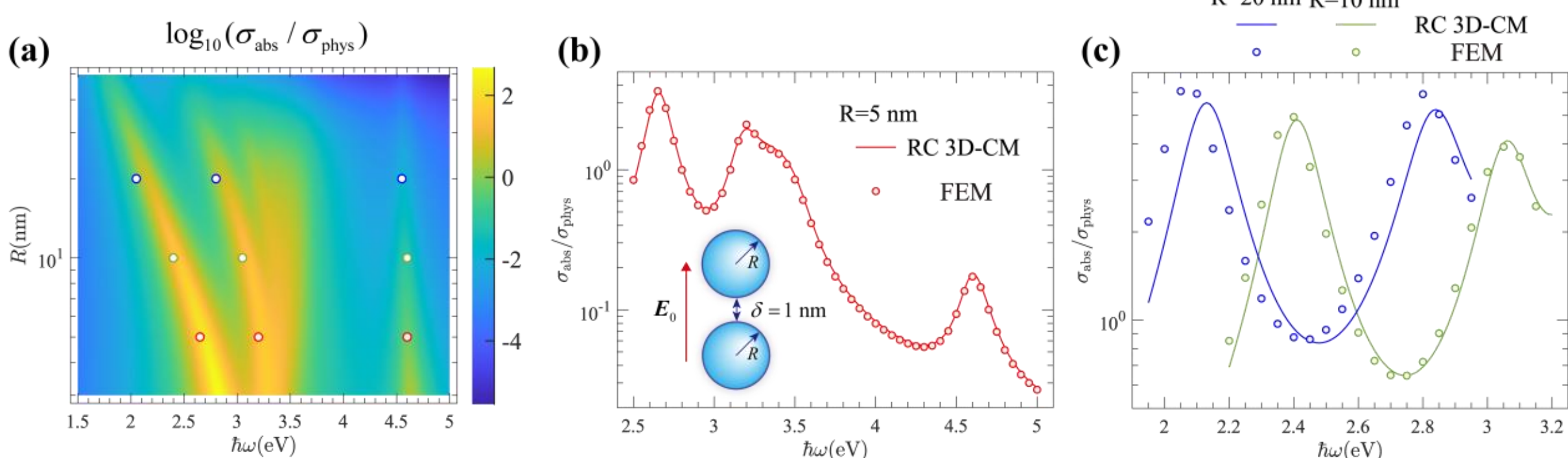

**Figure S6.** (a) Absorption spectra $\sigma_{\text{abs}}$ from the RC 3D-CM method, illustrated as a function of sphere radius $R$ and frequency $\omega$ for bi-sphere configuration with a fixed gap $\delta = 1$ nm. Open circles mark the LSP resonances calculated from FEM at $R = 5$ nm, 10 nm, and 20 nm. (b-c) Absorption spectra $\sigma_{\text{abs}}$ as a function of frequency $\omega$ at $R = 5$ nm (red), 10 nm (green), and 20 nm (blue). The solid lines and open circles are calculated by the RC 3D-CM method and FEM, respectively. The inset of (b) shows the calculation setup.

## Section VI. Implementation of response matrix in Lifshitz formula

Casimir energy of two bodies can be computed using a mode summation approach. This approach involves the summation of the zero-point energies $\frac{1}{2}\hbar\omega_i$ of eigenmodes in the physical system, which is renormalized by subtracting those of individual systems. For the two-body system, the complex eigenfrequencies are determined by the zeros of $f_m^{\text{c}} = \det[\mathbf{S}_{m,\text{wd}}]$, while for two single bodies, the complex eigenfrequencies are determined by the zeros of $f_m^{\text{s}} = \det\left[\mathbf{S}_{m,\text{wd}}^{(0)}\right]$, where $\mathbf{S}_{m,\text{wd}}^{(0)} = \begin{pmatrix} \mathbf{0} & {\mathbf{S}_{m,\text{wd}}^{++}}^{-1}\mathbf{T}_{m,\text{wd}}^{+-} \\ {\mathbf{S}_{m,\text{wd}}^{--}}^{-1}\mathbf{T}_{m,\text{wd}}^{-+} & \mathbf{0} \end{pmatrix}$ is response matrix of the decoupled system. Therefore, the mode condition function is expressed as

$$\begin{aligned} f_m(\omega) &= \frac{f_m^{\text{c}}}{f_m^{\text{s}}} \\ &= \det\left[{\mathbf{S}_{m,\text{wd}}^{(0)}}^{-1}\mathbf{S}_{m,\text{wd}}\right] \\ &= \det\left[\mathbf{I} - \left({\mathbf{T}_{m,\text{wd}}^{+-}}^{-1}\mathbf{S}_{m,\text{wd}}^{++}\right)\left({\mathbf{T}_{m,\text{wd}}^{-+}}^{-1}\mathbf{S}_{m,\text{wd}}^{--}\right)\right]. \end{aligned} \tag{S6.1}$$

Employing Cathy's argument principle [16], we obtain the Lifshitz formula as

$$E = \frac{1}{2\pi i} \sum_{m=0}^{+\infty} \int_{+i\infty}^{-i\infty} \frac{1}{2} \hbar\omega \frac{\partial}{\partial\omega} \ln f_m(\omega)\, \mathrm{d}\omega = \frac{\hbar}{4\pi} \sum_{m=0}^{+\infty} \int_{-\infty}^{+\infty} \ln f_m(i\xi)\, \mathrm{d}\xi\,. \tag{S6.2}$$

where $\omega = i\xi$.

## Section VII. Casimir softening

In Sec. VII A, we commence by deriving the Casimir softening functional $L_\eta$ in the reduction factor for two flat surfaces from the perspective of imaginary frequencies. Subsequently, in Sec. VII B, we analyze the contribution of Casimir softening through the spectral density of $L_\eta$. Finally, in Sec. VII C, we supplement the derivation of $L_\eta$ from real frequencies.

### A. Casimir softening functional using imaginary frequency approach

For two semi-infinite metal plates with permittivity $\epsilon_m = \epsilon_b - \frac{\omega_p^2}{\omega(\omega+i\gamma)}$ separated by $L$ [right panel in Fig. 3(a)], the Casimir force can be expressed as $F = \eta_F F_{\mathrm{PEC}}$ in terms of a reduction factor $\eta_F$, which measures a relative reduction in the Casimir force with respect to the case of perfect electric conductors (PECs). It is known that the reduction factor for the Casimir force is expressed as [17,18]

$$\eta_F = \frac{240}{\pi^3} \frac{L}{\lambda_p} \int_0^\infty K^2 dK \int_0^\infty d\Omega \sum_{a=s,p} \frac{r_a^2}{e^{2K_z} - r_a^2}, \tag{S7.1}$$

where $K = qL$ and $\Omega = \frac{\xi}{\omega_p}$ are dimensionless parallel wavevector and frequency, $K_z = \sqrt{K^2 + \Omega^2 \frac{L^2}{\lambda_p^2}}$, and $r_{s,p}$ denotes the reflection coefficients for two polarization modes (TE and TM, namely *s*- and *p*-polarizations in the main text). The reflection coefficients with *d*-parameters at the metal-vacuum interface are [19]

$$r_s^{\mathrm{wd}}[d_\sigma] = \frac{k_z - k_{z,\mathrm{m}} + (\epsilon_{\mathrm{m}} - 1) i k_0^2 d_\parallel}{k_z + k_{z,\mathrm{m}} - (\epsilon_{\mathrm{m}} - 1) i k_0^2 d_\parallel}, \tag{S7.2a}$$

$$r_p^{\mathrm{wd}}[d_\sigma] = \frac{\epsilon_{\mathrm{m}} k_z - k_{z,\mathrm{m}} + (\epsilon_{\mathrm{m}} - 1)\big(i q^2 d_\perp - i k_z k_{z,\mathrm{m}} d_\parallel\big)}{\epsilon_{\mathrm{m}} k_z + k_{z,\mathrm{m}} - (\epsilon_{\mathrm{m}} - 1)\big(i q^2 d_\perp + i k_z k_{z,\mathrm{m}} d_\parallel\big)}, \tag{S7.2b}$$

where $k_z = \sqrt{k_0^2 - q^2}$ and $k_{z,\mathrm{m}} = \sqrt{\epsilon_{\mathrm{m}} k_0^2 - q^2}$ are normal components of wavevectors in the vacuum and metal. When $d_{\perp,\parallel} = 0$, $r_{s,p}^{\mathrm{wd}}$ reduces to the classical reflection coefficients $r_{s,p}^{\mathrm{cl}} = r_{s,p}^{\mathrm{wd}}[d_\sigma = 0]$.

We start with the classical case, i.e., inserting $r_{s,p} = r_{s,p}^{\mathrm{cl}}$ into Eq. (S7.1). In the short distance $L \ll \lambda_p$ ($\lambda_p = 2\pi c/\omega_p$), the contribution to Casimir force mainly comes from the TM modes, and the reduction factor is expanded to the first order of $\frac{L}{\lambda_p}$

$$\eta_F = \alpha \frac{L}{\lambda_p}, \tag{S7.3}$$

where the coefficient

$$\alpha = \frac{240}{\pi^3} \int_0^\infty K^2 dK \int_0^\infty d\Omega \frac{1}{e^{2K}\left(r_p^{\mathrm{cl}}\right)^{-2} - 1} \tag{S7.4}$$

and

$$r_p^{\mathrm{cl}} = \frac{\epsilon_m - 1}{\epsilon_m + 1} = \frac{(\epsilon_b - 1)\Omega\left(\Omega + \gamma_p\right) + 1}{(\epsilon_b + 1)\Omega\left(\Omega + \gamma_p\right) + 1}, \tag{S7.5}$$

where $\gamma_\mathrm{p} = \frac{\gamma}{\omega_p}$. The coefficient $\alpha$ is 1.193 for a non-dissipative Drude metal ($\epsilon_b = 1$ and $\gamma = 0$), and the dissipation will decrease the value of $\alpha$ [17].

Next, we consider the mesoscopic case. Equation (S7.3) is rewritten as

$$\eta_F = \alpha \frac{L_\eta}{\lambda_p}, \tag{S7.6}$$

where $L_\eta$ is a functional on $d_\sigma$, associated with Casimir softening in the reduction factor. Hence, in the short distance ($L \ll \lambda_p$),

$$L_\eta = L \int_0^\infty dK \int_0^\infty d\Omega\, g[K, \Omega, d_\sigma(i\xi), L], \tag{S7.7}$$

where $g = \frac{240}{\pi^3 \alpha} \frac{K^2}{e^{2K}\left(r_p^{\mathrm{wd}}\right)^{-2} - 1}$ is the spectral density and the mesoscopic reflection coefficient is [20]

$$\begin{aligned} r_p^{\mathrm{wd}} &= \frac{\epsilon_m - 1 + (\epsilon_m - 1)(qd_\perp + qd_\parallel)}{\epsilon_m + 1 - (\epsilon_m - 1)(qd_\perp - qd_\parallel)} \\ &= \frac{(\epsilon_b - 1)\Omega\left(\Omega + \gamma_p\right) + 1 + Kd_\perp/L + Kd_\parallel/L}{(\epsilon_b + 1)\Omega\left(\Omega + \gamma_p\right) + 1 - Kd_\perp/L + Kd_\parallel/L}. \end{aligned} \tag{S7.8}$$

Furthermore, at the nanoscale, where $\frac{d_\sigma(i\xi)}{L} \ll 1$, Eq. (S7.7) can be expanded up to the first order $\mathcal{O}\left(\frac{d_{\perp,\parallel}(i\xi)}{L}\right)$ as

$$L_\eta\left[d_{\perp,\parallel}(i\xi); \omega_p, L\right] = L + \{L_\perp[d_\perp(i\xi)] + L_\parallel[d_\parallel(i\xi)]\}, \tag{S7.9}$$

where

$$L_{\perp}[d_{\perp}] = \frac{120}{\pi^3 \alpha} \int_0^\infty K^3 dK \int_0^\infty d\Omega \; \frac{1}{(\epsilon_b - 1)\Omega(\Omega + \gamma_p) + 1} d_{\perp} \begin{bmatrix} F_- - F_+ \\ +(e^K + 1)F_-^2 \\ -(e^K - 1)F_+^2 \end{bmatrix}, \quad \text{(S7.10)}$$

$$L_{\parallel}[d_{\parallel}] = \frac{120}{\pi^3 \alpha} \int_0^\infty K^3 dK \int_0^\infty d\Omega \; \frac{1}{(\epsilon_b - 1)\Omega(\Omega + \gamma_p) + 1} d_{\parallel} \begin{bmatrix} F_- - F_+ \\ -(e^K - 1)F_-^2 \\ +(e^K + 1)F_+^2 \end{bmatrix}, \quad \text{(S7.11)}$$

where

$$F_{\pm}(K, \Omega, \gamma_p) = \frac{1}{e^{K \frac{(\epsilon_b + 1)\Omega(\Omega + \gamma_p) + 1}{(\epsilon_b - 1)\Omega(\Omega + \gamma_p) + 1}} \pm 1}. \quad \text{(S7.12)}$$

If we assume non-dispersive $d_\sigma$, Eq. (S7.9) can be further rewritten as

$$L_\eta(d_\sigma, L) = L + \sum_{\sigma = \perp, \parallel} C_\sigma d_\sigma \,. \quad \text{(S7.13)}$$

Notably, if $\epsilon_b = 1$ and $\gamma_p = 0$, by integrating over $\Omega$ in Eqs. (S7.10)-(S7.11), we obtain

$$\begin{aligned} C_\perp = &\frac{30}{\pi^2} \int_0^\infty dK K^3 e^{-\frac{3}{4}K} \left( \frac{1}{\sqrt{\sinh \frac{K}{2}}} - \frac{1}{\sqrt{\cosh \frac{K}{2}}} \right) \\ &+ \frac{15}{\pi^2} \int_0^\infty dK K^3 e^{-\frac{3}{4}K} \left[ \frac{\cosh \frac{K}{2}}{\left(\sinh \frac{K}{2}\right)^{\frac{3}{2}}} - \frac{\sinh \frac{K}{2}}{\left(\cosh \frac{K}{2}\right)^{\frac{3}{2}}} \right], \end{aligned} \quad \text{(S7.14)}$$

$$C_\parallel = \frac{15}{\pi^2} \int_0^\infty dK K^3 e^{-\frac{3}{4}K} \left( \frac{1}{\sqrt{\sinh \frac{K}{2}}} - \frac{1}{\sqrt{\cosh \frac{K}{2}}} \right). \quad \text{(S7.15)}$$

Clearly, both coefficients no longer depend on $\omega_p$, and numerical calculations show $C_\perp = 5.2931$ and $C_\parallel = 0.708$. For various metals with distinct facets described by the permittivity in Sec. V B, these two coefficients are $C_\perp = 5.2727$ and $C_\parallel = 0.7273$ for Na, $C_\perp = 3.1610$ and $C_\parallel = 0.3793$ for Ag(100), and $C_\perp = 3.2207$ and $C_\parallel = 0.3876$ for Ag(111).

Moreover, we prove that $C_\sigma$ is positive definite. Given that $(K, \Omega, \gamma_p) > 0$ and $\epsilon_b(i\xi) - 1 \geq 0$, it is evident that $0 < e^K - 1 < e^K + 1$, $0 < F_+ < F_-$, and $1 - F_+(e^K + 1) < 1 - F_-(e^K - 1)$. Ultimately, we deduce $F_+ + (1 - e^K)F_+^2 < F_- + (1 + e^K)F_-^2$, and $F_+ - F_+^2(e^K + 1) < F_- - F_-^2(e^K - 1)$. Therefore, owing to the positive integral domain, $C_\sigma > 0$.

### B. Spectral density

To further elucidate the contribution from QSRs, we show the spectral density $g[K, \Omega, d_\sigma(i\xi)]$ as a function of wave vector and frequency in Figs. S7(a)-(b). Figure S7(a) is the classical scenario ($d_\sigma = 0$), demonstrating that primary correction to the classical attractive Casimir force of a Drude metal arises from evanescent waves at low frequencies and near $0.8\, L^{-1}$ wavevector. Meanwhile, the contribution from QSRs to Casimir softening happens at low frequencies and around $1.2\, L^{-1}$ wavevector, as depicted in Fig. S7(b).

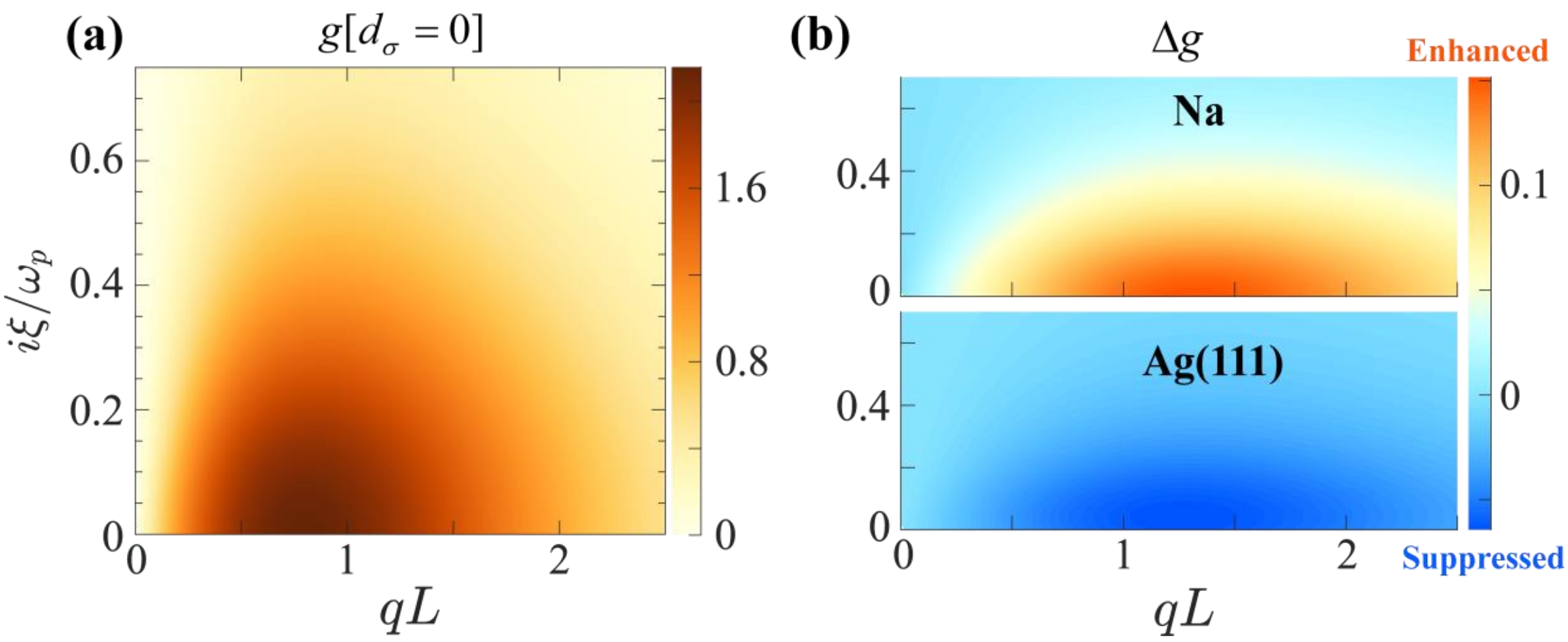

**Figure S7**. (a) Contour plot of $g[K, \Omega, d_\sigma = 0, L]$. (b) Spectral density difference $\Delta g(K, \Omega) = g[K, \Omega, d_\sigma, L] - g[K, \Omega, d_\sigma = 0, L]$ for Na and Ag(111). In (a-b), $L = 2$ nm.

### C. Casimir softening functional using real frequency mode sum

The Casimir energy can be expressed in terms of eigenfrequencies as

$$\frac{E}{A} = \frac{\hbar}{2} \sum_{a=\pm} \int_0^\infty \frac{q dq}{2\pi} (\omega_a - \omega_0), \tag{S7.16}$$

where $\omega_a$ and $\omega_0$ are the dispersions of two-coupling plates and a single infinite plate with $d_\sigma$ included, namely [21]

$$\omega_\pm = \frac{\omega_p}{\sqrt{2}} \sqrt{1 - qd_\perp + qd_\parallel \pm e^{-qL}(1 + qd_\perp + qd_\parallel)}, \tag{S7.17}$$

$$\omega_0 = \frac{\omega_p}{\sqrt{2}} \sqrt{1 - qd_\perp + qd_\parallel}. \tag{S7.18}$$

Notably, the system is assumed to be lossless, $\gamma_p = 0$. The Casimir force is then written as

$$\frac{F}{A} = \frac{\partial E}{A \partial L} = \frac{\hbar\, \omega_p}{2\sqrt{2}} \sum_{a=\pm} \int_0^\infty \frac{q^2 dq}{2\pi} \frac{a e^{-qL}(1 + qd_\perp + qd_\parallel)}{2\sqrt{1 - qd_\perp + qd_\parallel + a e^{-qL}(1 + qd_\perp + qd_\parallel)}}. \tag{S7.19}$$

Hence, the reduction factor is

$$\eta_F = -\frac{60}{\sqrt{2}\pi^2}\frac{L}{\lambda_p}\sum_{a=\pm}\int_0^\infty K^2 dK \frac{ae^{-K}\left(1+K\frac{d_\perp}{L}+K\frac{d_\parallel}{L}\right)}{2\sqrt{1-K\frac{d_\perp}{L}+K\frac{d_\parallel}{L}+ae^{-K}\left(1+K\frac{d_\perp}{L}+K\frac{d_\parallel}{L}\right)}} \quad (S7.20)$$

Similar to Eq. (S7.7), the equation above is expanded up to the first order $\mathcal{O}\left(\frac{d_{\perp,\parallel}(i\xi)}{L}\right)$ under the conditions of non-dispersive $d_\sigma$, allowing us to recover Eqs. (S7.13)-(S7.15) in the case $\epsilon_b = 1$.

**Section VIII. Nonlocal models in the framework of *d*-parameters**

The *d*-parameter formalism is undoubtedly beyond the HDM formalism, and thus, our recipe straightforwardly recovers the results based on the HDM and nonlocal models, which will be demonstrated next.

We start with the bi-parallel-plate configurations, which have been investigated in Refs. [22] and [23] by the nonlocal models. The calculations performed in these two papers are based on the Lifshitz formula with reflection coefficients corrected by the nonlocal models, i.e., Eqs. (11) and (14) in Ref. [22]. Our approach is logically the same and thus can recover the results therein. The Lifshitz formula Eq. (S7.1), which is the starting point of the Casimir softening framework, is exactly Eq. (11) in Ref. [22]. By setting $d_\perp = d_\perp^{\mathrm{HDM}} \equiv -\frac{i}{k_L} = -\frac{\beta}{\sqrt{\omega_p^2+\beta^2 Q^2-\omega(\omega+i\gamma)}}$ and $d_\parallel = d_\parallel^{\mathrm{HDM}} \equiv 0$ ( $k_L$ is the surface normal wave vector of longitudinal modes), our *d*-parameter-corrected reflection coefficients $r_{s,p}^{\mathrm{wd}}$ Eq. (S7.2) recover Eq. (14) in Ref. [22], and the ensuing results are essentially the same. Therefore, Refs. [22] and [23] can be treated as special cases in our treatments.

However, as mentioned previously, the *d*-parameter formalism is beyond the HDM formalism and nonlocal models. To demonstrate, we calculate the QSR correction factor $\Xi_{\mathrm{QSR}}$ for the bi-parallel-plate configuration using both $d_\perp^{\mathrm{HDM}}$ and our used *d*-parameters [Fig. 2(a)-(b) in the main text], as shown in Fig. S8. The HDM formalism always predicts a suppression in the nanoscale Casimir force (red, cyan, and blue dashed lines in Fig. S8), whatever the metals are (Fig. 2 in Ref. [22]). The physical reason is that the induced surface electrons always spill inside the bulk due to the naturally hard-wall assumptions at the surface within the HDM formalism. However, it is known that the induced surface electrons can spill in or out of the bulk depending on metals and crystal facets. Therefore, its influence on the nanoscale Casimir force should be distinct, as revealed by our method. The solid lines in Fig. S8 show $\Xi_{\mathrm{QSR}}$ calculated by our method, and a suppression (enhancement) in the nanoscale Casimir force has been seen in the noble metal Ag (alkali metal Na). Such a difference cannot be seen from the HDM formalism and nonlocal models.

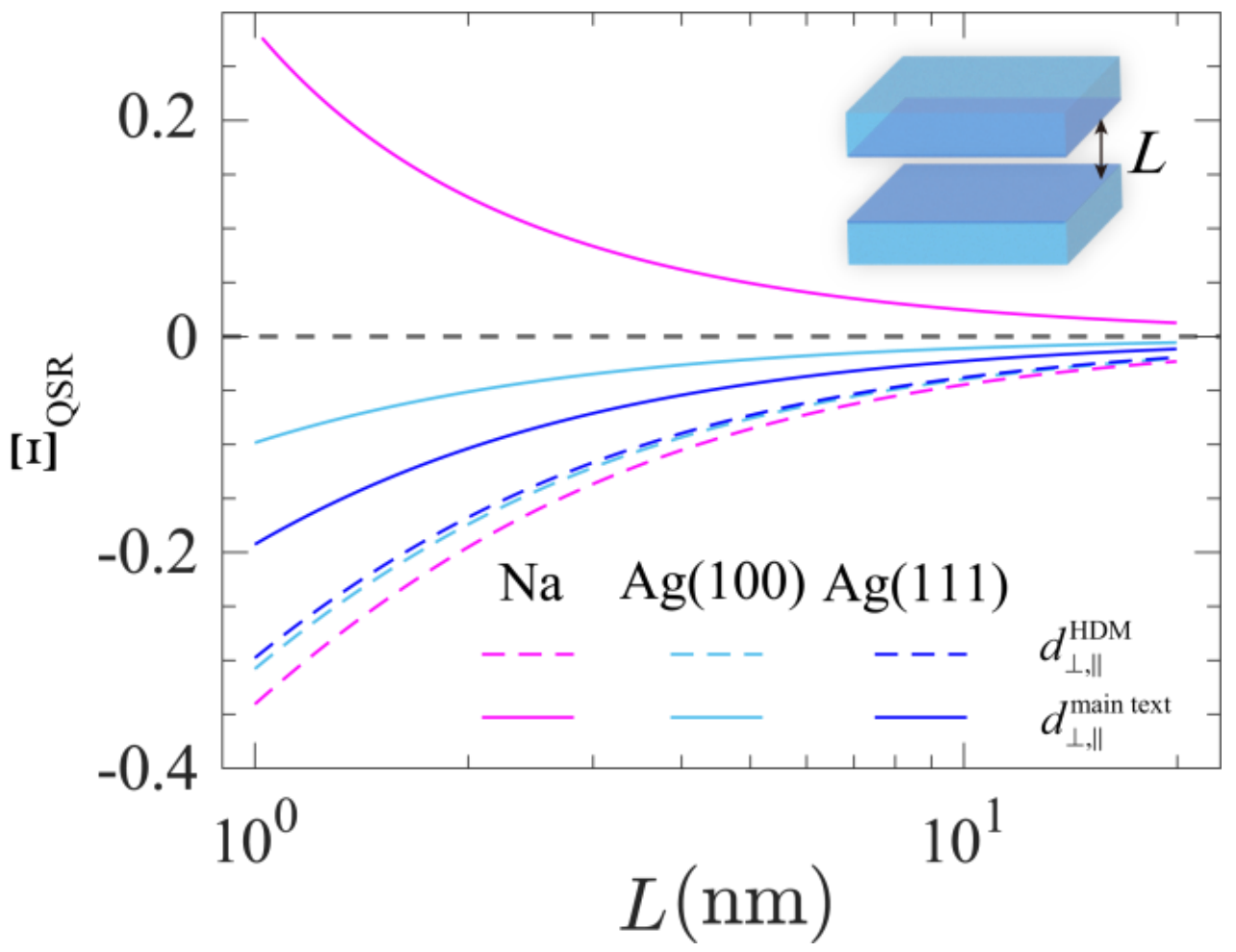

**Figure S8.** QSR correction factor $\Xi_{\mathrm{QSR}}$ as a function of distance $L$ for the bi-parallel-plate configuration. The solid lines (dashed lines) are calculated by using $d_{\perp,\parallel}$ in the main text ($d_{\perp,\parallel}^{\mathrm{HDM}}$ corresponded to Ref. [22]) for Na (red), Ag(100) (cyan), and Ag(111) (blue).

Second of all, Ref. [24] by P. T. Kristensen *et al*. has established a high-accuracy time domain method to investigate the Casimir-Polder interaction between a molecule and other micro-structured materials. The method is excellent, and it is straightforward to include the nonlocal models. For the atom-surface configuration, the Casimir-Polder interaction is calculated by [25]

$$E(L) = -\frac{\hbar}{2\pi}\mathrm{Im}\int_0^{+i\infty} d\omega\, \boldsymbol{\alpha}(\omega)\cdot\boldsymbol{G}\left(L,\omega;r_{s,p}\right), \tag{S8.1}$$

where $\boldsymbol{\alpha}(\omega) = 4\pi\varepsilon_0\alpha(\omega)\boldsymbol{I}$ is isotropic polarizability tensor and $\boldsymbol{G}\left(L,\omega;r_{s,p}\right) = \frac{1}{8\pi\varepsilon_0}\int_0^\infty qdq\,\kappa\left[\left(r_{s,p}+\frac{\omega^2}{c^2\kappa^2}r_{s,p}\right)\left[\hat{\boldsymbol{e}}_x\hat{\boldsymbol{e}}_x+\hat{\boldsymbol{e}}_y\hat{\boldsymbol{e}}_y\right]+2\frac{q^2}{\kappa^2}r_{s,p}\hat{\boldsymbol{e}}_z\hat{\boldsymbol{e}}_z\right]e^{-2\kappa L}$ is Green tensor dependent on reflection coefficients $r_{s,p}$. The inclusion of nonlocal models in Ref. [24] is by substituting $r_{s,p}$ with $r_{s,p}^{\mathrm{HDM}}$, which can be obtained by setting $d_\perp = d_\perp^{\mathrm{HDM}}$ and $d_\parallel = d_\parallel^{\mathrm{HDM}}$ in our $d$-parameter-corrected reflection coefficients Eq. (S7.2). Within our $d$-parameter formalism, we have used the $d$-parameter-corrected reflection coefficients $r_{s,p}^{\mathrm{wd}}$ in the Green tensor $\boldsymbol{G}$. For the noble metals, for example, Au and Ag, $r_{s,p}^{\mathrm{wd}}$ and $r_{s,p}^{\mathrm{HDM}}$ are qualitatively similar, and thus, our results, say Fig. 2(d) in our main text, are qualitatively the same as the results shown in Fig. 7 of Ref. [24].

Lastly, the *d*-parameters formalism can also recover the nonlocal Casimir force results in Ref. [26] by J. Sun *et al*. The key lies in how the *d*-parameter-corrected scattering coefficients recover the nonlocal scattering coefficients, say Eqs. (9)-(10) in Ref. [26], for a system composed of metal nanoparticles embedded in a homogenous permittivity background. For such a system, the *d*-parameter-corrected scattering coefficients have been given by Eqs. 4(a)-(b) in Ref. [19]. When $d_\perp = -\frac{j_l(x_L)}{x_L j_l'(x_L)} \frac{\epsilon_d(\epsilon_m-\epsilon_\infty)}{\epsilon_\infty(\epsilon_m-\epsilon_d)} R$ and $d_\parallel = 0$ ($x_L = k_L R$), the *d*-parameter-corrected scattering coefficients completely can recover the nonlocal scattering coefficients Eqs. (9)-(10) in Ref. [26] for a non-magnetic case. From the absorption and extinction spectra point of view, our method can predict the same conclusion as Ref. [19] for a single sphere, and thus, our work recovers it in the nanoscale.

**Section IX. Retardation effect in the *d*-parameter formalism**

Whether the retardation effects significantly embody at the nanoscale or not depends on the physical quantity under consideration. Within the scale of interest in this work, the influence of retardation effects on the *d*-parameter corrections to the nanoscale Casimir force is minor, and the details are in the following.

Take the planar surfaces as the first example. Figure S9(a) depicts a comparison between retarded (red) and nonretarded (quasi-static, blue) dispersions for a single Na surface (dashed lines) and the ensuing bi-parallel-plate configuration (solid lines). Note that all of them have been calculated using the *d*-parameters employed in our work. Whatever the single surface or coupled surface case, retardation effects are indeed remarkable when the parallel wavenumber $q$ is comparable to or larger than $\omega/c$. This is undoubtedly the typical $q$ region governing the near-field interaction between nanoscale objects. However, when $q$ further increases to the magnitude of Fermi wavenumber $k_F$, the retardation effects become ignorable. Since Casimir energy is the sum of zero-point energy [Eq. (S7.16)], namely, the dispersion in Fig. S9(a), the contribution from the retardation effect to Casimir energy in the nanoscale occurs in a portion of $q$ space.

However, the *d*-parameters do affect the dispersion within all the $q$ ranges, whether the retardation effect has been taken into account or not. The gray and black lines in Fig. S9(a) show the dispersions in the classical limit (without *d*-parameters), and their deviation compared to those with *d*-parameters included is observable and dramatic, whatever the retardations. It is then expected that the $d$-parameter corrections are dominant at the nanoscale whether the retardation effect is included or not.

To examine the impact of retardation effects on the nanoscale Casimir force, we compare the retarded and non-retarded Casimir forces for both Na and Ag(111) scenarios, as depicted in Fig. S9(b). We see that the retardation correction is positively correlated to the distance $L$, but the $d$-parameter correction, as demonstrated in the main text, is inversely correlated to $L$, whatever the retardations. To quantify, we plot in Fig. S9(c) the $d$-parameter corrections for both Na and Ag with (solid lines) and without (circles) retardations. It is clear that at the nanoscale of interest here, the $d$-parameter corrections are present and predominant regardless of retardation. The influence of retardation effects on the $d$-parameter corrections is minor, as seen from Fig. S9(c).

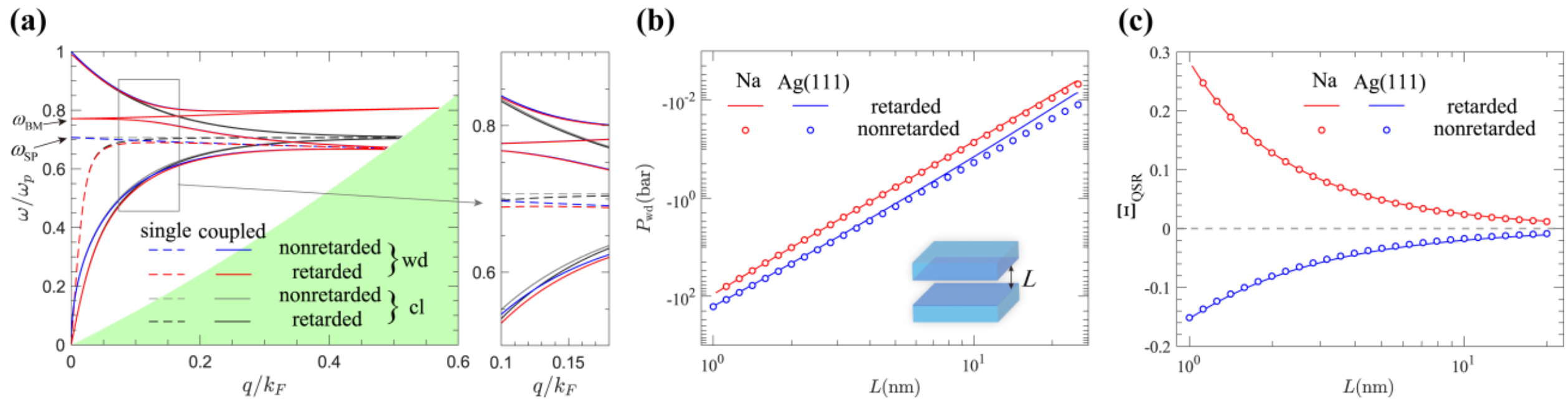

**Figure S9.** (a) Retarded (red) and nonretarded (blue) dispersions with $d$-parameters (*wd*) and retarded (black) and nonretarded (gray) dispersions without $d$-parameters (*cl*). The dashed lines show the result of a single Na surface, while the solid lines show the result of two coupled Na surfaces separated by a distance. The light-green hatched area is the bulk electron-hole pair region. A local zoom is shown on the right side. In the $d$-parameter framework, for the single surface case, there are two dispersive curves, i.e., surface plasmon polariton (SPP, $\omega \sim \omega_p/\sqrt{2}$) and Bennett mode (BM, $\omega \sim 0.78\,\omega_p$), while for the coupled surface case, there are four dispersive curves. (b) Casimir force per unit area $P_{wd}$ as a function of distance $L$ for the bi-parallel-plate configuration. (c) QSR correction factor $\Xi_{QSR}$ as a function of distance $L$. In (b-c), solid lines show retarded results, while circles are nonretarded results for Na case (red colors) and Ag(111) case (blue colors).

The above conclusion can be generalized beyond the planar surfaces, and we next estimate the retardation effects in both the bi-sphere and sphere-plate configurations. The retardation-corrected formula is $F_r = F_{nr} - F_{dipole,nr} + F_{dipole,r}$, where $F_r$ and $F_{nr}$ are retarded and non-retarded Casimir forces for sphere-plate or bi-sphere configurations, and $F_{\mathrm{dipole,r}}$ and $F_{\mathrm{dipole,nr}}$ are retarded and non-retarded Casimir forces for dipole-plate (atom-surface) or dipole-dipole configurations. We can now estimate the retardation effects on the $d$-parameter corrections in the Casimir force for both the bi-sphere [Fig. S10(a)] and sphere-plate [Fig. S10(b)] configuration. Although the results are distinct from the bi-parallel-plate configuration, the conclusions are similar, i.e., the $d$-parameter corrections are still present and predominant regardless of retardation, as shown in Fig. S10(c-d).

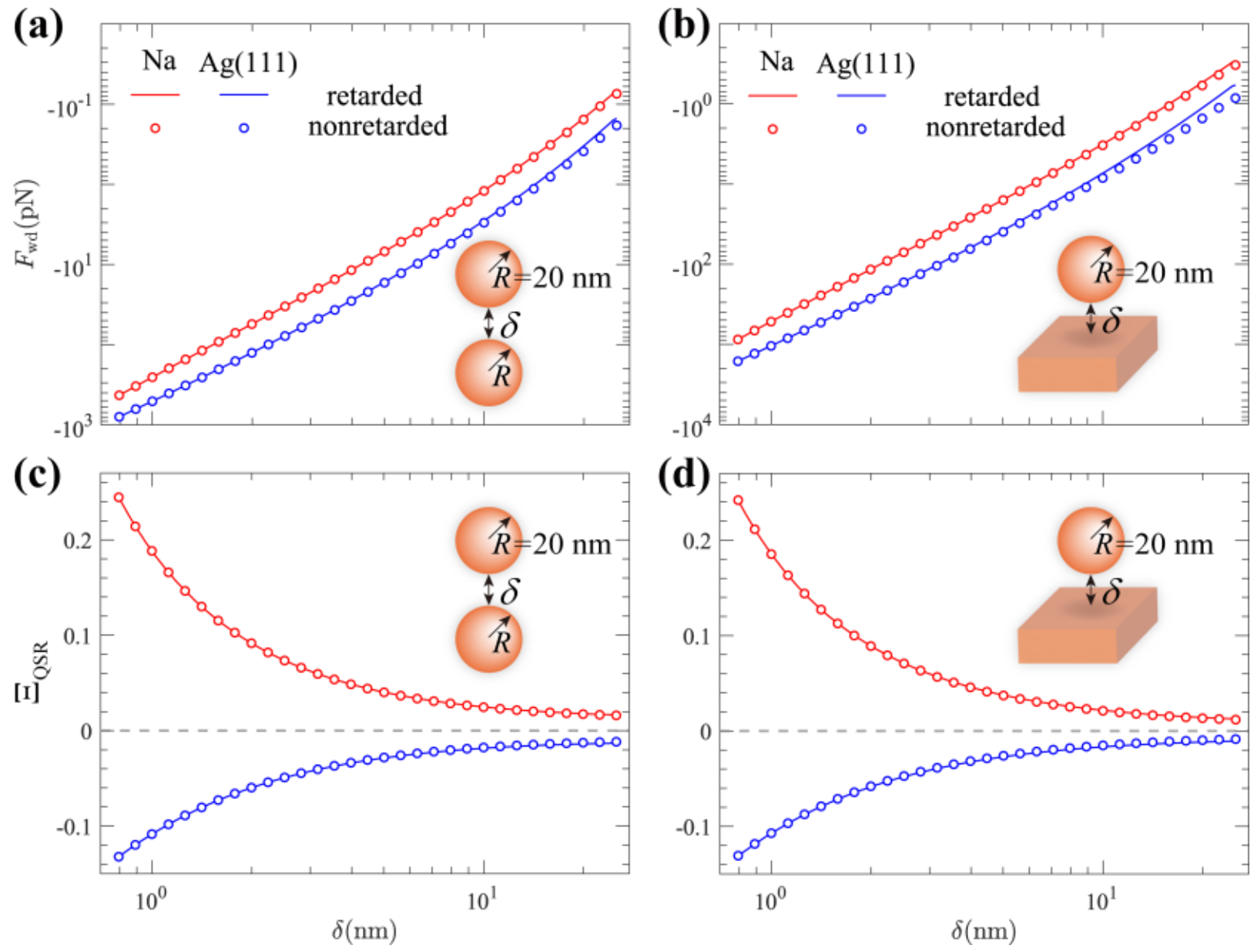

**Figure S10.** (a-b) Retardation-corrected Casimir force $F_{\mathrm{wd}}$ as a function of gap $\delta$ for a bi-sphere (sphere-plate) configuration with two (one) fixed sphere radii $R = 20$ nm, as shown in the inset. (c-d) QSR correction factor $\Xi_{\mathrm{QSR}}$ as a function of gap $\delta$. In (a-d), solid lines show retarded results, while circles show nonretarded results for Na case (red colors) and Ag(111) case (blue colors).

### Section X. Reference image surface method

The reference image plane method [27] can be related to our approach with some generalizations, and the role of Casimir softening distance in our approach is similar to the reference plane position in the reference image plane method. We detail the relations in the following.

#### A. Failure of the reference image plane treatments to the nanoscale objects

The reference image plane treatment to the atom-surface configuration is depicted in Fig. S11(a). Since the electrostatic problem for any fluctuation charges can be handled by the method of images, people in the 80's have employed such a method. However, the image plane needs to be shifted from the crystalline planar plane because the centroid of induced electron density is not precisely at the crystalline plane. To showcase such a recipe, we calculate the van der Waals force by imposing the following reference image plane position for the metal-vacuum interface [28]

$$d_{\mathrm{IP}}(\omega = i\xi) = \frac{d_{\parallel}(\omega = i\xi) + \epsilon_m(\omega = i\xi) d_{\perp}(\omega = i\xi)}{\epsilon_m(\omega = i\xi) + 1}, \tag{S10.1}$$

where $\epsilon_m$ is the permittivity of the metal. The results of $d_{\mathrm{IP}}$ are shown by the red (real frequency) and blue (imaginary frequency) lines in Fig. S11(b), and the van der Waals force by using $d_{\mathrm{IP}}$ is shown by the red solid line in Fig. S11(c). For comparison, we also calculate it through Eq. (S8.1) with the $d$-parameter-corrected reflection coefficients $r_{s,p}^{\mathrm{wd}}$, and the results are shown by the red open circles in Fig. S11(c). The good agreement reveals the validity of the reference image plane treatment. It is worth pointing out that $d_{\mathrm{IP}}(\omega)$ depends on frequency, and thus, the reference image plane position varies during the calculation.

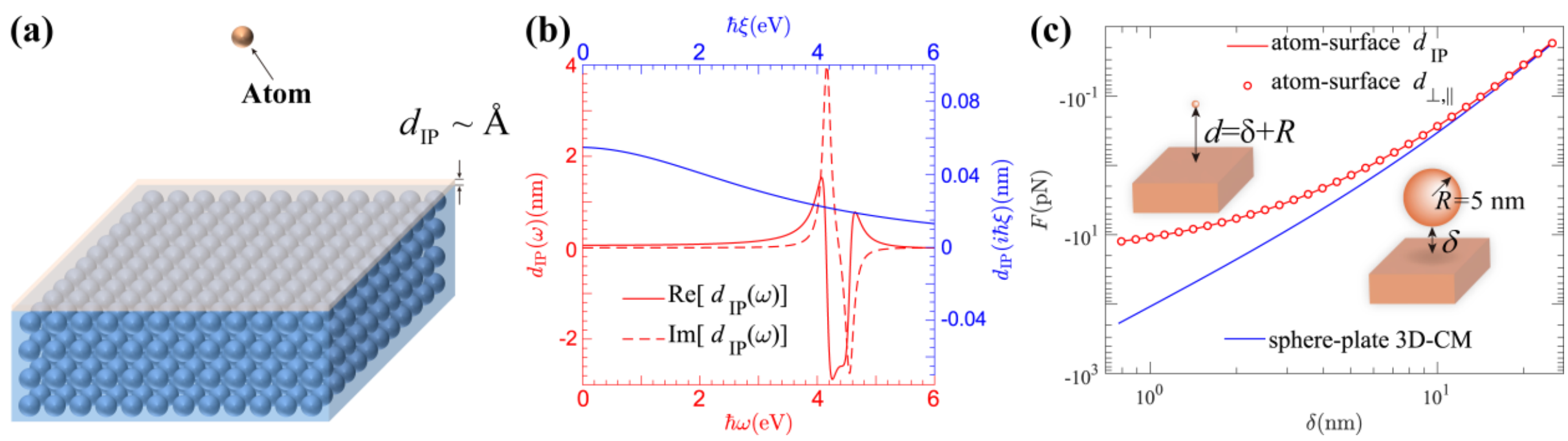

**Figure S11.** (a) Schematics of the reference image plane method for an atom-surface configuration. (b) Reference image plane position $d_{\mathrm{IP}}(\omega)$ $[d_{\mathrm{IP}}(\omega = i\xi)]$ as a function of real frequency $\omega$ ($i\xi$). The red (blue) lines show $d_{\mathrm{IP}}(\omega)$ $[d_{\mathrm{IP}}(\omega = i\xi)]$ results. (c) Casimir force $F$ as a function of gap $\delta$. The red solid line and circles represent the results from the reference image plane method and $d$-parameter formalism [Eq. (S8.1) with $d$-parameters] for an atom-surface configuration, while the blue solid line represents the results for sphere-plate configuration with a fixed sphere radius $R = 5$ nm.

We now turn to the sphere-plate configuration [the bottom right inset in Fig. S11(c)], and the results calculated by the 3D-CM method are shown by the blue solid line in Fig. 11(c). We deliberately choose the sphere to make its effective dipolar polarizability the same as the one used by the red solid line. It is seen that the dipole approximation is correct when $\delta$ is over $20$ nm and fails when $\delta$ is within the scale of interest here because of the near-field nature. This indicates that when investigating the van der Waals/Casimir interaction between objects with nanoscale separations, we cannot simply replace the objects with dipoles, hindering the usage of image planes. However, our focus is indeed to study the van der Waals/Casimir interaction between complex objects that are favorable experimentally [the bottom right inset in Fig. S11(c) and also Fig. 1(a) in the main text], and thus, the reference image plane treatment cannot directly apply.

**B. Reference image surface method: generalization of reference image plane treatment**

To generalize the reference image plane idea and correlate it with our recipe, we begin with the Casimir force between two parallel plates, i.e., Eq. (5) in the main text. We expand the *d*-parameter-corrected *p*-polarized reflection coefficient [20] to its first order in $q$

$$r_p^{\mathrm{wd}} = \frac{\epsilon_m - 1 + (\epsilon_m - 1)(qd_\perp + qd_\parallel)}{\epsilon_m + 1 - (\epsilon_m - 1)(qd_\perp - qd_\parallel)} \approx r_p^{\mathrm{cl}}(1 + 2qd_{\mathrm{IP}}) \approx r_p^{\mathrm{cl}} e^{2qd_{\mathrm{IP}}}. \tag{S10.2}$$

Substituting the above expansion into Eq. (5) in our main text, we can obtain the following Casimir force expression

$$F(L) = -\frac{\hbar}{2\pi^2}\int_0^\infty \mathrm{d}\xi \int_0^\infty q^2 \mathrm{d}q \frac{\left(r_p^{\mathrm{cl}}\right)^2 e^{-2q(L-2d_{\mathrm{IP}})}}{1-\left(r_p^{\mathrm{cl}}\right)^2 e^{-2q(L-2d_{\mathrm{IP}})}}. \tag{S10.3}$$

Since $L$ is the distance between two parallel plates, the above equation indicates that the Casimir force is now equivalent to the scenario of two parallel plates separated by $L - 2d_{\mathrm{IP}}(\omega = i\xi)$ but with the classical reflection coefficients. In other words, each plate experiences a frequency-dependent shift $d_{\mathrm{IP}}(\omega = i\xi)$, which undoubtedly generalizes the reference image plane idea to the bi-parallel-plate configuration and paves the way to complex objects under consideration in this work.

Since PFA is a straightforward approach to handling complex objects with a not-bad and acceptable accuracy, we utilize the PFA to push the reference image plane idea further. Take Fig. S12(a) for an example. Due to the pairwise addition nature of PFA, we propose to shift each object surface to their reference image positions given by $d_{\mathrm{IP}}(\omega = i\xi)$, perform the classical electromagnetic calculations at each frequency, and finally, perform the Lifshtiz integral to have the Casimir force. We term this approach as the reference image surface (RIS) method. Concretely to Fig. S12(a), the spherical surface and planar surface, respectively, shift $d_{1,\,\mathrm{IP}}(\omega)$ and $d_{2,\,\mathrm{IP}}(\omega)$, leading $\delta$ effectively changing to $\delta - d_{1,\,\mathrm{IP}}(\omega) - d_{2,\,\mathrm{IP}}(\omega)$. Figure S12(b) shows the QSR correction factor $\Xi_{\mathrm{QSR}}$ calculated using the 3D-CM (solid lines) and RIS (open circles) methods. The excellent agreement demonstrates the validity of the RIS method, which delivers a recipe to handle complex objects.

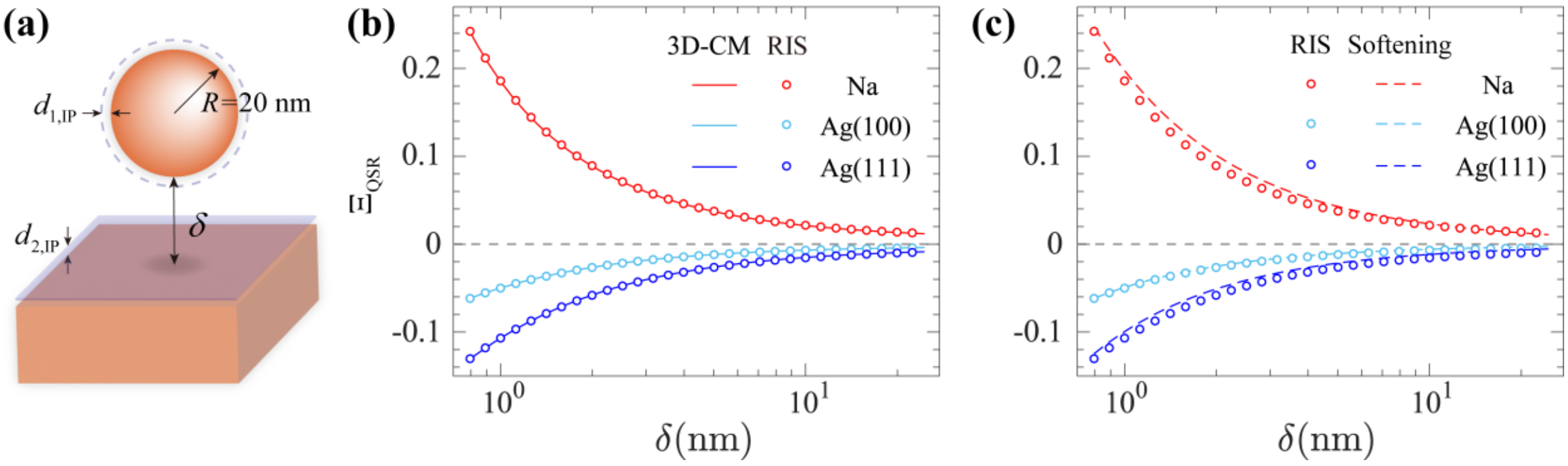

**Figure S12.** (a) Schematic depicting the RIS method for the sphere-plate configuration. (b-c) QSR correction factor $\Xi_{\mathrm{QSR}}$ as a function of gap $\delta$, calculated using the 3D-CM (solid lines), RIS (open circles), and the generalized Casimir softening [PFA combined with Eq. (6) in the main text] (dashed lines) methods. The Na, Ag(100), and Ag(111) results are represented by the red, cyan, and blue colors, respectively. The radius is chosen to $R = 20$ nm.

### C. Relation to the Casimir softening distance

The utilization of the RIS method shown in Fig. S12(a,b), besides its merit, suffers several drawbacks compared with the spirit of PFA. As shown in Fig. S11(b), the values of $d_{\mathrm{IP}}(\omega)$ vary with frequency, making the RIS method still numerically not that straightforward because one remains to perform the Lifshitz integral. Moreover, the sign of $d_{\mathrm{IP}}(\omega)$ is possible to change with real frequency (red lines), indicating that the image surface can be either inside or outside the crystalline surface and thus impede tracing the nanoscale Casimir force simply by checking the effective distance.

To resolve this conundrum, we recall the reference plane position $Z_0$ defined as associated with the reference image plane position $d_{\mathrm{IP}}(\omega = i\xi)$ [28]. In contrast to the dispersive characteristics of $d_{\mathrm{IP}}(\omega = i\xi)$, $Z_0 = \frac{1}{4\pi C}\int_0^\infty d\xi\ \alpha \frac{\epsilon_m - 1}{\epsilon_m - 1}$ ($C$ is the van der Waals constant of atom-surface [29]) is the physical distance, which merely depends on the materials and surfaces, and can be directly plugged into the van der Waals interaction formula for the atom-surface configuration as

$$E(Z) = -\frac{C}{(Z - Z_0)^3}. \tag{S10.4}$$

Crucially, the adoption of $Z_0$ not only avoids the Lifshitz integral but also makes the enhancement and suppression of the nanoscale Casimir force transparent. By recognizing such merits, we then should work out a similar quantity in the RIS method, which is nothing but what the Casimir softening formula [Eq. (6) in the main text] aims for. To see this, we express the nanoscale Casimir force between two parallel plates as

$$F(L) = -\frac{D}{\left(L - L_{Z_0}\right)^3}, \qquad D = \frac{\pi^2 \hbar c \alpha}{240 \lambda_p}, \tag{S10.5}$$

where $D$ is the Casimir constant of two parallel plates and $L_{Z_0} = \sum C_\sigma d_\sigma / 3$. The quantity $\sum C_\sigma d_\sigma /3$ in Eq. (6) plays the same role as $Z_0$ because they share the same properties. Further extending to the sphere-plate geometry, there is a good agreement between the results from RIS and PFA combined with Eq. (6) methods, as shown in Fig. S12(c). This highlights our Casimir softening recipe in two aspects, namely, the enhancement and suppression mechanism of the nanoscale Casimir force becomes transparent, and the investigation of the Casimir force between complex objects is achievable under the PFA.

**Section XI. Bi-sphere system**

For the nanoscale gap, our physical explanation extends beyond the sphere-plate system to encompass the bi-sphere configuration. We maintain the radius of one sphere while progressively altering the shape of another object from a flat plate to a small sphere, as depicted in the inset in Fig. S13(a). The qualitative consistency of the QSR correction factor persists, as shown in Fig. S13(a). Moreover, for the bi-sphere setup, we further confirm the accuracy of our 3D-CM method by employing the multipole expansion method (MEM) in conjunction with Feibelman $d$-parameters for calculating the Casimir energy as

$$E = \frac{\hbar}{4\pi i} \sum_{m=0}^{+\infty} \int_{-\infty}^{+\infty} \ln \det[\mathbf{I} - \mathbf{M}\mathbf{R_1}\mathbf{N}\mathbf{R_2}] \, \mathrm{d}\omega \,, \tag{S11.1}$$

where $M_{l,l'} = (-1)^{l'+m} \frac{(l+l')!}{(l-m)!(l'+m)!}$ and $N_{l,l'} = (-1)^{|l-l'|} M_{l,l'}$ are associated with transformations between the spherical basis at different centers and $R_{1(2),l,l'} = \delta_{l,l'} \frac{(\epsilon_m - \epsilon_d)\left[1 + \frac{l}{R}\left(d_\perp + \frac{l+1}{l} d_\parallel\right)\right]}{\epsilon_m + \frac{l+1}{l}\epsilon_d - (\epsilon_m - \epsilon_d)\frac{l+1}{R}(d_\perp - d_\parallel)} \left(\frac{R_{1(2)}}{R_1 + R_2 + \delta}\right)^{2l+1}$ relates to the multipole polarizability of the sphere [19].

Figure S13(b) demonstrates excellent agreement between both methods. (Note that MEM is not available for the sphere-plate system.) The inset of Fig. S13(b) shows that under the same convergence tolerance, our 3D-CM method requires a lower truncation order $l_c$ compared to MEM, especially when $\delta \ll R$, implying more computationally efficient. For the large $\frac{\delta}{R}$ case, both methods require similarly low $l_c$, consequently leading to comparable computational efficiency. Therefore, our 3D-CM method not only accurately calculates both systems but also demonstrates more efficient convergence.

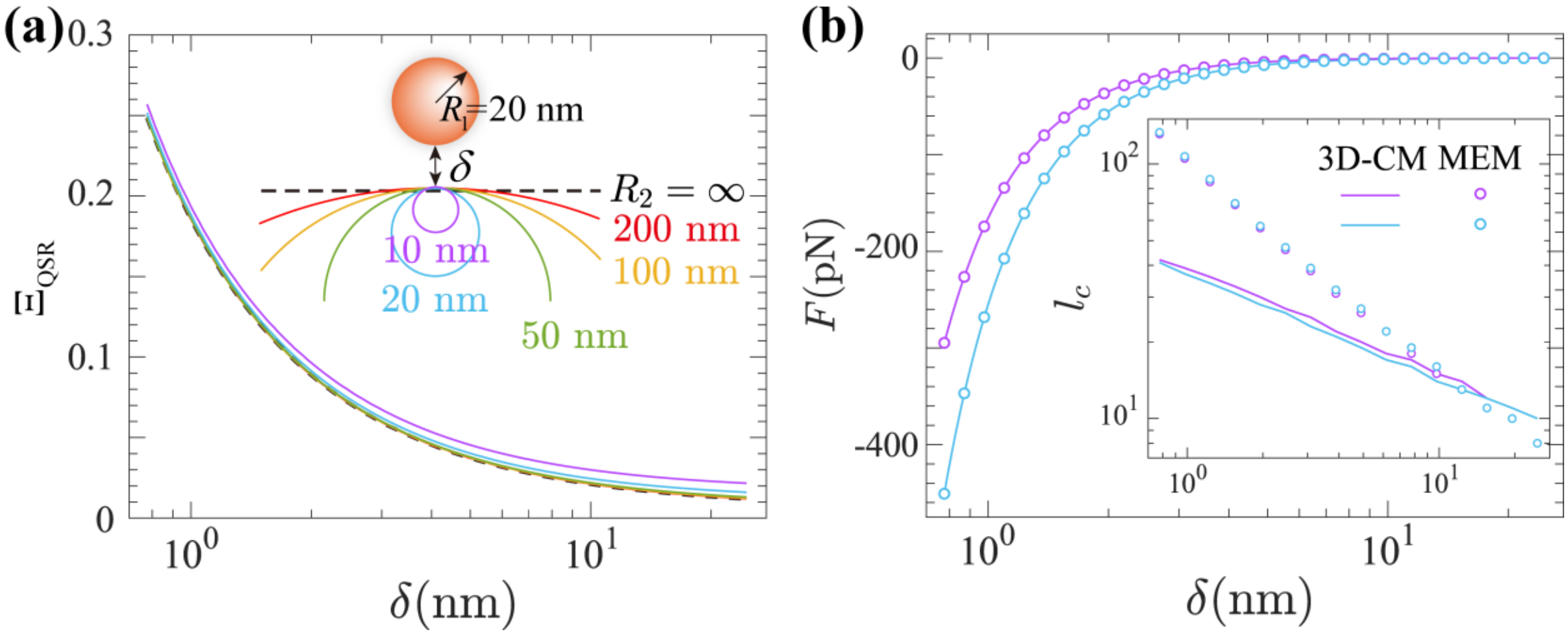

**Figure S13**. (a) $\Xi_{\mathrm{QSR}}$ of the sodium bi-sphere system, consisting of one sphere with $R_1 = 20\ \mathrm{nm}$ and another sphere with its radius decreasing from $R_2 = \infty$ to $R_2 = 10\ \mathrm{nm}$. (b) Comparison of Casimir force calculated by the 3D-CM (solid lines) and MEM (open circles). The inset shows the minimum truncation order $l_c$ needed for both methods to achieve a relative accuracy of $10^{-5}$.

To further reveal our recipe, we also calculate the Casimir forces for the bi-sphere system by using RIS and generalized softening methods, as shown in Fig. S14, demonstrating a good agreement with the results from the 3D-CM method. This indicates that the Casimir softening framework developed above for the sphere-plate system is still valid in digesting the role of various metal and crystal facets in the bi-sphere system.

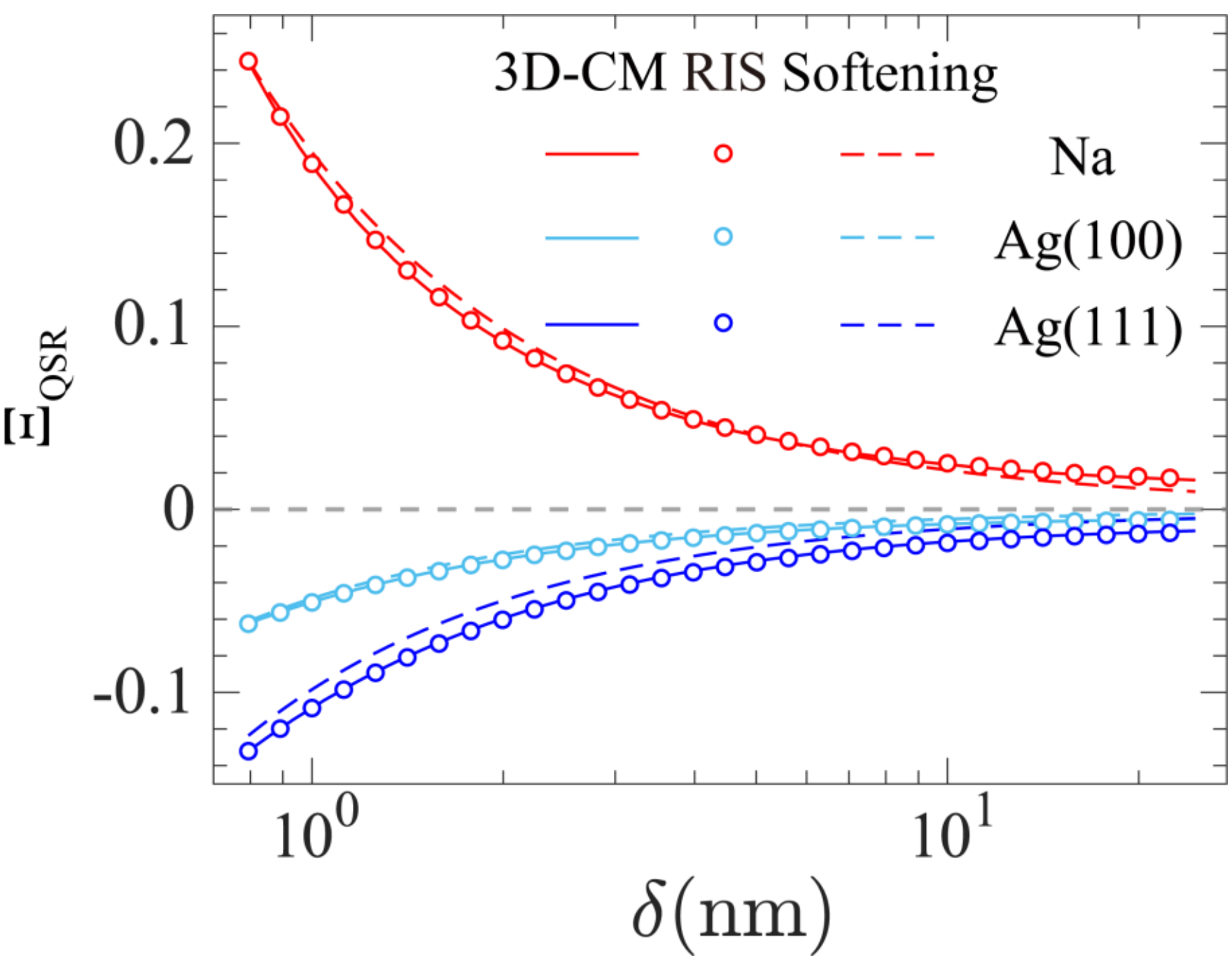

**Figure S14.** $\Xi_{\mathrm{QSR}}$ of the bi-sphere system with $R_1 = R_2 = 20$ nm as a function of gap $\delta$, calculated using the 3D-CM (solid lines), RIS (open circles), and the generalized Casimir softening [PFA plus Eq. (6) in the main text] (dashed lines) methods. The Na, Ag(100), and Ag(111) results are represented by the red, cyan, and blue colors, respectively.

**Section XII. Impact of nanoscale Casimir softening on the power-law exponents**

For the sphere-plate configuration, the classical Casimir force $F_{\mathrm{cl}}$ exhibits an inverse quadratic dependence on the gap size $\delta$ at short distances, i.e., $F_{\mathrm{cl}} \propto \delta^{-2}$. The power-law exponent, such as -2, is a significant physical quantity in experimental measurements. However, at nanoscale distances, Casimir softening affects this exponent. To quantify such impact, we fit numerical data obtained from our 3D-CM by

$$F = c\delta^n, \tag{S12.1}$$

where $c$ and $n$ are the coefficient and power-law exponent independent of $\delta$. Figure S15 shows the fitting results on a double-logarithmic scale for the Na, Ag(100), and Ag(111) cases, demonstrating excellent agreement with the 3D-CM results at small gap sizes. The extracted power-law exponents listed in Table S3 indicate that Casimir softening decreases the exponent by approximately 10% for Na, while increasing it for the Ag cases (up to 5%). Additionally, these quantitative changes persist even when using a modified power-law expression $F = c(\delta - \delta_0)^n$ to fit the 3D-CM results, where $\delta_0$ accounts for a calibration offset related to experimental positioning.

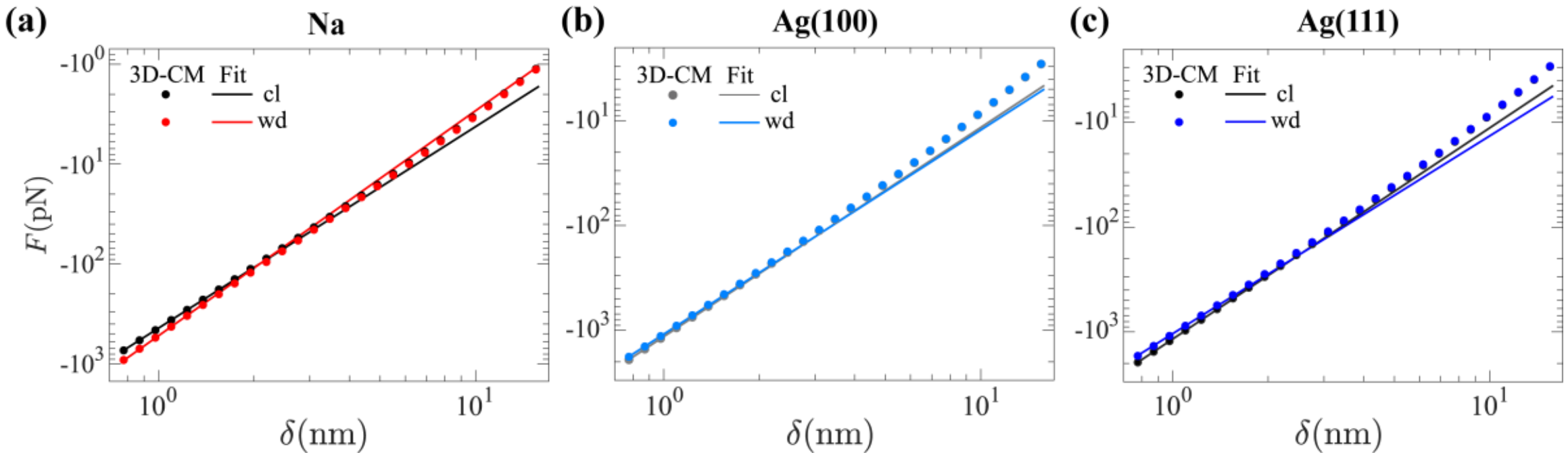

**Figure S15.** (a-c) Casimir force of the sphere-plate system as a function of $\delta$ for (a) Na, (b) Ag(100), and (c) Ag(111). The filled dots (lines) represent our 3D-CM calculation results (power law fitting results). The black dots and lines denote the classical (cl) results, while the red, cyan, and blue ones show the results with $d$-parameters (wd) included. The sphere radius is chosen to $R = 20$ nm.

**Table S3.** Power-law exponents $n_{\mathrm{cl}}$ and $n_{\mathrm{wd}}$. The last row shows $\Delta n = n_{\mathrm{wd}} - n_{\mathrm{cl}}$. Note that the trust-region algorithm is employed, and the fitted values are featured with at least 95% confidence.

| | Na | Ag(100) | Ag(111) |
|---|---|---|---|
| $n_{\mathrm{cl}}$ | -2.018 | -1.998 | -2.017 |
| $n_{\mathrm{wd}}$ | -2.254 | -1.949 | -1.889 |
| $\Delta n$ | -0.236 | 0.049 | 0.128 |

## References

[1] J. B. Pendry, A. I. Fernández-Domínguez, Y. Luo, and R. Zhao, Nat. Phys. **9**, 518-522 (2013).

[2] R. Zhao, Y. Luo, A. I. Fernandez-Dominguez, and J. B. Pendry, Phys. Rev. Lett. **111**, 033602 (2013).

[3] M. T. H. Reid, A. W. Rodriguez, J. White, and S. G. Johnson, Phys. Rev. Lett. **103**, 040401 (2009).

[4] Y. Yang, D. Zhu, W. Yan, A. Agarwal, M. Zheng, J. D. Joannopoulos, P. Lalanne, T. Christensen, K. K. Berggren, and M. Soljačić, Nature **576**, 248-252 (2019).

[5] D. E. Blair, *Inversion Theory and Conformal Mapping* (American Mathematical Society, U. S. A., 2000).

[6] U. Leonhardt and T. G. Philbin, in *Progress in Optics*, Vol. 53 (Elsevier, New York, 2009), pp. 69-152.

[7] N. B. Kundtz, D. R. Smith, and J. B. Pendry, Proc. IEEE **99**, 1622-1633 (2011).

[8] J. B. Pendry, D. Schurig, and D. R. Smith, Science **312**, 1780-1782 (2006).

[9] D. Schurig, J. J. Mock, B. J. Justice, S. A. Cummer, J. B. Pendry, A. F. Starr, and D. R. Smith, Science **314**, 977-980 (2006).

[10] M. Rahm, S. A. Cummer, D. Schurig, J. B. Pendry, and D. R. Smith, Phys. Rev. Lett. **100**, 063903 (2008).

[11] L. D. Landau, L. P. Pitaevskii, and E. M. Lifshitz, *Electrodynamics of Continuous Media* (Pergamon Press, Oxford, 1984).

[12] A. Rodríguez Echarri, P. A. D. Gonçalves, C. Tserkezis, F. J. García de Abajo, N. A. Mortensen, and J. D. Cox, Optica **8**, 710-721 (2021).

[13] F. Yang and K. Ding, Phys. Rev. B **105**, L121410 (2022).

[14] P. B. Johnson and R. W. Christy, Phys. Rev. B **6**, 4370-4379 (1972).

[15] E. D. Palik, *Handbook of Optical Constants of Solids II* (Academic, NewYork, 1991).

[16] M. Bordaga, U. Mohideenb, and V. M. Mostepanenko, Physics Reports **353**, 1 (2001).

[17] A. Lambrechta and S. Reynaud, Eur. Phys. J. Plus **8**, 309–318 (2000).

[18] F. Intravaia, C. Henkel, and A. Lambrecht, Phys. Rev. A **76**, 033820 (2007).

[19] P. A. D. Gonçalves, T. Christensen, N. Rivera, A.-P. Jauho, N. A. Mortensen, and M. Soljačić, Nat. Commun. **11**, 366 (2020).

[20] P. J. Feibelman, Prog. Surf. Sci. **12**, 287-407 (1982).

[21] T. Christensen, W. Yan, A. P. Jauho, M. Soljacic, and N. A. Mortensen, Phys. Rev. Lett. **118**, 157402 (2017).

[22] R. Esquivel, C. Villarreal, and W. L. Mochán, Phys. Rev. A **68**, 052103 (2003).

[23] R. Esquivel, C. Villarreal, and W. L. Mochán, Phys. Rev. A **71**, 029904(E) (2005).

[24] P. T. Kristensen, B. Beverungen, F. Intravaia, and K. Busch, Phys. Rev. B **108**, 205424 (2023).

[25] W. Beiglböck, *Casimir Physics* (Springer, Heidelberg, 2011).

[26] J. Sun, Y. Huang, and L. Gao, Phys. Rev. A **89**, 012508 (2014).

[27] B. N. J. Persson and P. Apell, Phys. Rev. B **27**, 6058-6065 (1983).

[28] B. N. J. Persson and E. Zaremba, Phys. Rev. B **30**, 5669-5679 (1984).

[29] J. N. Israelachvili, *Intermolecular and surface forces* (Elsevier, California, 2011).